\documentclass[conference]{IEEEtran}

\usepackage[T1]{fontenc}
\usepackage{xspace}
\usepackage{subfig}

\newcommand{\theTitle}{TLS in the wild: An Internet-wide analysis of TLS-based protocols for electronic communication}

\newif\ifstatus
\statusfalse

\usepackage[firstinits=true,sortcites,backend=bibtex]{biblatex}
\usepackage{graphicx}
\usepackage{booktabs}
\usepackage[group-digits=true, group-separator={,}]{siunitx}

\newcommand{\zmap}{\texttt{zmap}\xspace}
\newcommand{\starttls}{STARTTLS\xspace}
\newcommand{\tls}{TLS\xspace}
\newcommand{\ssltls}{SSL/TLS\xspace}
\newcommand{\ssl}{SSL~3\xspace}
\newcommand{\oldssl}{SSL~2\xspace}
\newcommand{\ocsp}{OCSP\xspace}
\newcommand{\pitm}{MitM\xspace}
\newcommand{\latinlocution}[1]{\textit{#1}}
\newcommand{\etc}{etc.\xspace}
\newcommand{\eg}{\latinlocution{e.g.,}\xspace}
\newcommand{\ie}{\latinlocution{i.e.,}\xspace}
\newcommand{\etal}{\latinlocution{et al.}\xspace}
\newcommand{\ctos}{client-to-server\xspace}
\newcommand{\stos}{server-to-server\xspace}

\begin{document}

\title{\theTitle}
\author{
    \IEEEauthorblockN{Ralph Holz\IEEEauthorrefmark{1}\thanks{$^{*}$The work was carried out during the first author's time at Data61/CSIRO.}, Johanna Amann\IEEEauthorrefmark{3}, Olivier Mehani\IEEEauthorrefmark{2}, Matthias Wachs\IEEEauthorrefmark{4}, Mohamed Ali Kaafar\IEEEauthorrefmark{2}}

  \IEEEauthorblockA{\IEEEauthorrefmark{1}University of Sydney, Australia, Email: {ralph.holz}@sydney.edu.au}
  \IEEEauthorblockA{\IEEEauthorrefmark{2}Data61/CSIRO, Sydney, Australia, Email: {name.surname}@data61.csiro.au}
  \IEEEauthorblockA{\IEEEauthorrefmark{3}ICSI, Berkeley, USA, Email: johanna@icir.org}
  \IEEEauthorblockA{\IEEEauthorrefmark{4}Technical University of Munich, Germany, Email: wachs@net.in.tum.de}
  \textbf{This is a preprint of the camera-ready version to appear at NDSS 2016. Last update: 19 Dec 2015.}
}
\IEEEoverridecommandlockouts
\makeatletter\def\@IEEEpubidpullup{9\baselineskip}\makeatother

\IEEEpubid{\parbox{\columnwidth}{Permission to freely reproduce all or part of
        this paper for noncommercial purposes is granted provided that copies
        bear this notice and the full citation on the first page. Reproduction
        for commercial purposes is strictly prohibited without the prior
        written consent of the Internet Society, the first-named author (for
        reproduction of an entire paper only), and the author's employer if the
        paper was prepared within the scope of employment.  \\ NDSS '16, 21-24
February 2016, San Diego, CA, USA\\ Copyright 2016 Internet Society, ISBN
1-891562-41-X\\ http://dx.doi.org/10.14722/ndss.2016.23055 }
\hspace{\columnsep}\makebox[\columnwidth]{}}

\maketitle

\begin{abstract}
  Email and chat still constitute the majority of electronic communication on the
Internet. The standardisation and acceptance of protocols such as SMTP, IMAP,
POP3, XMPP, and IRC has allowed to deploy servers for email and chat in a
decentralised and interoperable fashion. These protocols can be secured by
providing encryption with TLS---directly or via the STARTTLS extension. X.509
PKIs and ad hoc methods can be leveraged to authenticate communication peers.
However, secure configuration is not straight-forward and many combinations of
encryption and authentication mechanisms lead to insecure deployments and
potentially compromise of data in transit.
In this paper, we present the largest study to date that investigates the
security of our email and chat infrastructures.  We used active Internet-wide
scans to determine the amount of secure service deployments, and employed
passive monitoring to investigate to which degree user agents actually choose
secure mechanisms for their communication. We addressed both \ctos interactions
as well as \stos forwarding. Apart from the authentication and encryption
mechanisms that the investigated protocols offer on the transport layer, we
also investigated the methods for client authentication in use on the
application layer.  Our findings shed light on an insofar unexplored area of
the Internet.  Our results, in a nutshell, are a mix of both positive and
negative findings. While large providers offer good security for their users,
most of our communication is poorly secured in transit, with weaknesses in the
cryptographic setup and especially in the choice of authentication mechanisms.
We present a list of actionable changes to improve the situation.

\end{abstract}

\section{Introduction}
\label{sec:introduction}

Despite the rise of mobile messaging and some more centralised, newer
communication platforms, two forms of electronic, (nearly) instant messaging still
remain dominant on the public Internet: email and chat. Of the two, email is
the most pervasive form of communication ever, with over 4.1 billion accounts
in 2014, predicted to reach over 5.2 billion in 2018~\cite{Radicati}.  As for
chat, the most widely used standard-based networks are IRC group chats and the
XMPP instant messaging and multi-user conferencing network.

In their early days, email protocols such as SMTP, POP3, and IMAP were designed
with no special focus on security.  In particular, authentication in SMTP was
introduced a while after the protocol's standardisation, initially as a way to
fight spam. User agents started to move towards encryption and authenticated
connections gradually, using the then-new \ssl and later the \tls protocols to
protect the transport layer.  \ssltls can provide authentication, integrity,
and confidentiality.  Where \ssltls is not used, user credentials may be
transmitted in plaintext, with no protection against eavesdropping, and message
bodies can be tampered with (unless end-to-end mechanisms like OpenPGP or
S/MIME are used, which is a comparatively rare setup).

Although \ssltls support mutual authentication, the most common usage pattern
in the context of email and chat is unilateral authentication: only the
responder of a communication is authenticated on the transport layer.  The
primary reason for this is the protocols' reliance on an X.509 Public Key
Infrastructure (PKI) for authentication purposes\footnote{Variants of \tls that
support other forms of authentication have been standardised, but seem to be
rarely used.} and the subsequent need for client certification, an operation
that is expensive in practice, introduces much administrative overhead, and
often also requires user education.  In most cases, initiators are 
authenticated on the application layer instead, \ie by mechanisms that are
specific to the application layer protocol in question.  Passwords schemes are
the most common choice, although any mechanism that is supported by both initiator and
responder is possible.  Different password schemes offer
varying levels of security---\eg the password may be sent without further
protection over the \ssltls channel, or a challenge-response mechanism like
CRAM, or even SCRAM, may be used.  The latter is particularly elegant as it does
not require the responder to store the actual password, nor is the password ever
sent over the connection.  The choice of password scheme has a profound
influence on security in case of missing authentication on the level of
\ssltls.

The proper in-band authentication of the responder is a key element in \ssltls.
X.509 certificates are used for this purpose. These are issued by so-called
Certificate Authorities (CAs), which are trusted parties whose trust anchors
(so-called root certificates) are shipped with common software (\eg operating
systems, browsers, mail clients, \ldots).  Unfortunately, it is known today
that X.509 PKIs often suffer from poor deployment practices.
Holz~\etal~\cite{Holz2011} were the first to show this in a large-scale,
long-term study for the Web PKI.  Durumeric~\etal~\cite{Durumeric2013a} later
extended the study to all Internet hosts, confirming the earlier findings.
However, no such work exists for the electronic communication protocols on
which we rely every day.

In this paper, we present the largest measurement study to date that
investigates the security of \ssltls deployments for email and chat. Based on
our findings, we derive recommendations to achieve better overall security.  We
employ both active Internet-wide scans as well as passive monitoring. Active
scans are used to characterise global server installations, \ie how servers are
configured to act as responders in a \ssltls connection.  Passive monitoring
allows us to investigate the actual security parameters in use when initiators
establish \ssltls connections.

From 2015-06-30 through 2015-08-04, we actively scanned the IPv4 address space
(3.2B routable addresses), with one scanning run for each protocol we analysed.
We connected to the standard ports for the considered protocols: SMTP/STARTTLS,
SMTPS, SUBMISSION, IMAP/STARTTLS, IMAPS, POP3/STARTTLS, and POP3S for email;
for chat, we investigated IRC/STARTTLS, IRCS, XMPP/STARTTLS, and XMPPS. We
performed complete \ssltls handshakes. This allowed us to establish a list of
current deployments (a total of more than 50M active ports), and collect
certificates, cipher suite offers, and cryptographic parameters. Where
applicable, we also sent application-layer messages to request the list of
supported methods for authentication on the application layer.  Orthogonally to
this, nine days of passive monitoring (2015-07-29 through \mbox{2015-08-06}) of a link
serving more than \num{50000} users showed more than 16M connections to about
\num{14000} different services.  We captured the same set of \ssltls and
authentication-related parameters from this monitoring data as we did for
active scans. This allows us to compare usage by actual clients to the simple
existence of a deployed service.  As a reference and comparison point, we also
considered HTTPS and traffic on port 443 in both active and passive
measurements as the deployment of this protocol is particularly well
understood.

We analysed this data to evaluate the security of connections and deployments.
We considered the validity of the certificates and the practices of the issuing
CAs, the quality of cryptographic parameters, software, and \ssltls versions, as
well as the authentication methods.  In a nutshell, we have both negative and
positive findings to report. Considering active scans, we find that there is
much room for improvement. For example, for the IMAPS servers that
completed the \tls handshake, we report that just under 40\% also had correct certificate
chains deployed.  We found such low rates for all protocols---the
best-provisioned service was in fact SMTPS, where just over 40\% of servers had
valid certificate chains. SMTP/STARTTLS, which is used to forward emails
between mail exchange servers, showed a rate of just 30\%.  For chat, we found
the best results for XMPPS in \stos forwarding: 27\% of servers offered valid
certificates.

When considering data from our passive monitoring and investigating
\emph{connections} rather than server deployments, the situation seems much
better, at least at a first glance: the vast majority of connections is
encrypted and uses valid certificates (with SMTP/STARTTLS again showing poorer 
numbers, however).  This is due to the fact that large providers such as
Gmail or Hotmail are properly configured and offer good security, and most
connections go to these providers.  However, we also found that it is common
that the STARTTLS extension is not supported by servers that receive less
connections. In these cases, connections are not encrypted at all. This is
again particularly often true for email.  This phenomenon suggests a 
likelihood that communication is often not sufficiently secured in transit
between mail exchange servers, unless both sender and receiver are customers of
large providers.

The rest of this paper is organised as follows. The next section presents
background for \ssltls, PKI, and the studied protocols. It also gives an
overview of related work. We describe our data collection methods and datasets
in Section~\ref{sec:datacollection} and data analysis in
Section~\ref{sec:securityanalysis}. Based on our findings, we identify risks
and threats in Section~\ref{sec:risksthreats}. We suggest some pathways towards
improving the situation before concluding in Section~\ref{sec:conclusion}.

\section{Background and Related Work}
\label{sec:background}

\subsection{Standard messaging protocols}

The messaging protocols in common use today have been specified by the IETF
over the years; use with \ssltls or the STARTTLS extension was added later. For
example, the original RFC~821 for SMTP is from 1982, but the STARTTLS extension
for SMTP was specified in 1999.  Other protocols experienced similar organic
development, and the result is a variety of ways in which \ssltls is used in
email and chat.

\paragraph{Electronic Mail}

Email relies on two sets of protocols: one for email transfer and one for
retrieval. The Simple Mail Transfer Protocol (SMTP)~\cite{rfc5321} is the
cornerstone of email distribution systems. Its primary purpose was message
transfer: so-called Message Transfer Agents (MTAs) forward messages by
establishing an SMTP session to the next MTA on the path to the destination,
until they arrive at their final destination.  SMTP is also used as a
submission protocol\footnote{This was possibly first made explicit in
RFC~2476~\cite{rfc2476}.}: in a nutshell, user agents---\eg `email clients' such as
Thunderbird---submit mails from a local computer for further delivery to a `mail
server' that is commonly operated by the user's
service provider.  `Webmail' solutions such as GMail blur the distinctions
between mail submission and mail transfer somewhat: they offer web-hosted
front-ends for mail composition; mail submission and mail transfer are handled
entirely transparently on server-side.  SMTP was initially operated on port 25.
Later, port 587 was specified to be used for message submission~\cite{rfc6409}
by potentially authenticated submitters, in an attempt to differentiate between
legitimate activity and spam. Nevertheless, port 25 still remains in use for
both purposes, message transfer and submission.

Once at the destination server, email can be retrieved using either of two
protocols. The Post Office Protocol (version~3, commonly referred to as
POP3)~\cite{rfc1939} operates on port 110 and allows a remote client to
download newly-received emails to a local mailbox. The Internet Message Access
Protocol (version~4, commonly called IMAP)~\cite{rfc3501} uses port 143 and
offers access, manipulation, and download of messages in a mailbox stored on
the server side.

\paragraph{Chat and Instant Messaging}

Instant chat is an old concept, which predates even the Web. Internet Relay
Chat (IRC)~\cite{rfc1459} is a protocol that allows a number of IRC clients to
connect to an IRC server and join so-called channels (chat rooms) or have
private conversations. Messages, especially on channels, are relayed between
IRC servers. An oddity of IRC deployments is that server-to-server
communication is implementation-dependent (despite a specification
in~\cite{rfc2813}). Over time, this has led to IRC servers clustering into a
number of distinct `IRC networks'.  While the official IANA port for
client-to-server connections is 194, IRC is most commonly deployed on port
6667 instead~\cite{rfc7194}, but other ports are also sometimes used.  The
ports for \stos communication are specific to the IRC network.

In the footsteps of the proprietary instant messaging (IM) networks of the late
1990s, the more general XML-based eXtensible Messaging and Presence Protocol
(XMPP) was specified. Its core functionality is defined in
RFC~6120~\cite{rfc6120}, IM extensions in RFC~6121~\cite{rfc6121}.  Further
extensions exist.  Similar to the SMTP infrastructure, a number of XMPP servers
exchange messages on behalf of their users as part of the XMPP IM
network\footnote{XMPP Instant Messaging was known as Jabber before its
standardisation.}. The protocol uses port 5222 for \ctos communication, and
5269 for \stos forwarding.  XMPP, with or without proprietary extensions, is
also used in non-federated enterprise or proprietary services\footnote{E.g., HipChat
 uses a flavour of XMPP, as did the early Google Talk.}.

\subsection{\ssltls}

\tls 1.0 is the IETF-standardised version of \ssl\footnote{\ssl, originally created
at Netscape, was never standardised by the IETF, but later captured in a
historic RFC~\cite{rfc6101}.}. All versions before \tls 1.0, \ie \oldssl and
\ssl, are deprecated today. \tls is at version 1.2 and contains many critical
fixes that remove weaknesses of previous versions. Version 1.3 is currently in
the standardisation process. As \ssl and \tls 1.0 are very similar and a few
pockets of \ssl use remain, we speak of \ssltls when our findings apply to both
\ssl and \tls.  All email and chat protocols can be used with \ssltls. In IMAP
and POP3, only \ctos communication occurs. SMTP and XMPP define both \ctos and
\stos communication patterns. For IRC, once again only the \ctos pattern is
properly defined.

There are two ways to negotiate an \ssltls session. The first is to use \ssltls
directly. This requires a well-known port, \ie assignment of a new, dedicated
port by IANA. Application layer protocols that use this method are often
indicated by adding a 'S' at the end, \eg HTTPS, IMAPS, etc. Clients that do
not support \ssltls may still connect to the normal port. In \stos
communication, the servers may use certificates to authenticate to each other
(\ie either unilateral or mutual authentication may be used). As stated in the
introduction, in \ctos communication it is common that only the server is
authenticated; the client is authenticated later as part of the application
layer protocol.  In the case of SMTP, port 465 was initially defined for SMTPS,
but was deprecated later~\cite{draft-hoffman-smtp-ssl-04} in favour of
\starttls (see below). It is nevertheless still used.  The
dedicated ports for IMAPS and POP3S are 993 and 995, respectively.  For IRCS,
several exist~\cite{rfc7194}, with 6697 being very commonly used for client
connections.  XMPP does not have standard ports for \ssltls, but ports 5223 and
5270 are prevalent for \ctos and \stos communication, respectively.

The second major way to use \ssltls is to connect with TCP on the normal port
first and then upgrade the connection using a protocol-specific command.  This
method is commonly referred to as \starttls. The specifications in the RFCs
commonly require clients to first query a server for \starttls support with a
specific `capability' command before trying to upgrade the
connection~\cite{rfc2595}.  The server can confirm an upgrade; the \ssltls
handshake follows.  This is specified for SMTP (particularly for SUBMISSION) in
\cite{rfc3207}, in \cite{rfc2595} for IMAP and POP3, and in \cite{rfc7590} for
XMPP.  While \starttls is not formally specified for IRC, the InspIRCd
implementation\footnote{\texttt{https://wiki.inspircd.org/STARTTLS\_Documentation}}
is generally considered a reference.

\starttls has the advantage that no dedicated port is required and that the
communication partners can decide dynamically whether they want to use \ssltls.
A major limitation is the vulnerability to active \pitm attacks, where an
attacker interferes with the \starttls-related commands.  Unless clients or 
servers are specifically configured not to allow any connection without
upgrade, the attack succeeds and the entire communication will be in
plain.  Depending on the user agent, users may not even be prompted with a security
warning. The attack has been observed in the wild
\cite{2015durumeric_empirical_analysis_mail_security}.  

The \ssltls handshake is the same for both forms of connection establishment.
The initiator sends an initial message together with information which
symmetric cipher suites and \ssltls protocol versions it can support. The
responder picks a cipher suite and negotiates a protocol version in its reply.
It also sends an X.509 PKI certificate to authenticate. In another round trip,
the cryptographic parameters---which may also include Diffie--Hellman
parameters for forward secrecy---are then confirmed. The entities that wish to
authenticate also include a proof that they are in possession of the private
key that corresponds to their certificate.

It should be noted that email transfer over \ssltls is generally designed to
prioritise successful transfer over any security guarantees.  An opportunistic
approach to security is often favoured by implementations: both initiator and
responder may choose to ignore any authentication problems and proceed with
message delivery despite errors or warnings.

\subsection{X.509 PKI}

In order for an entity to have trust into the authentication step, a number of
conditions must be fulfilled that pertain to the configuration of the X.509 PKI
in use.  First and foremost, CAs must only issue certificates after applying
due diligence in identifying the party that wishes to be certified.  The
CA/Browser forum has established guidelines for the Web use
case~\cite{CABForumBR2011}.  The so-called Baseline Requirements define due
diligence to require at least an (usually automated) check if the requesting
party can receive email under the requested domain name and a specific email
address.\footnote{There are alternatives, \eg some form of token can be
    published on the web server, and some CAs apply further checks, \eg lookups
of WHOIS.} However, previous work has revealed cases where even this
basic diligence was neglected. These cases are documented in, \eg
\cite{holz_phd,Ristic2013}.  On several occasions, CAs have been compromised.
Since any CA may issue certificates for any domain, compromise of one CA is
enough to compromise the entire PKI. More details on relevant attacks on the
X.509 PKI for the Web can also be found in~\mbox{\cite{Ristic2013, holz_phd}}. Notably
however, X.509's use in email and chat protocols remains largely unexplored. 

\begin{figure}[tb]
  \centering
  \includegraphics[width=0.25\textwidth]{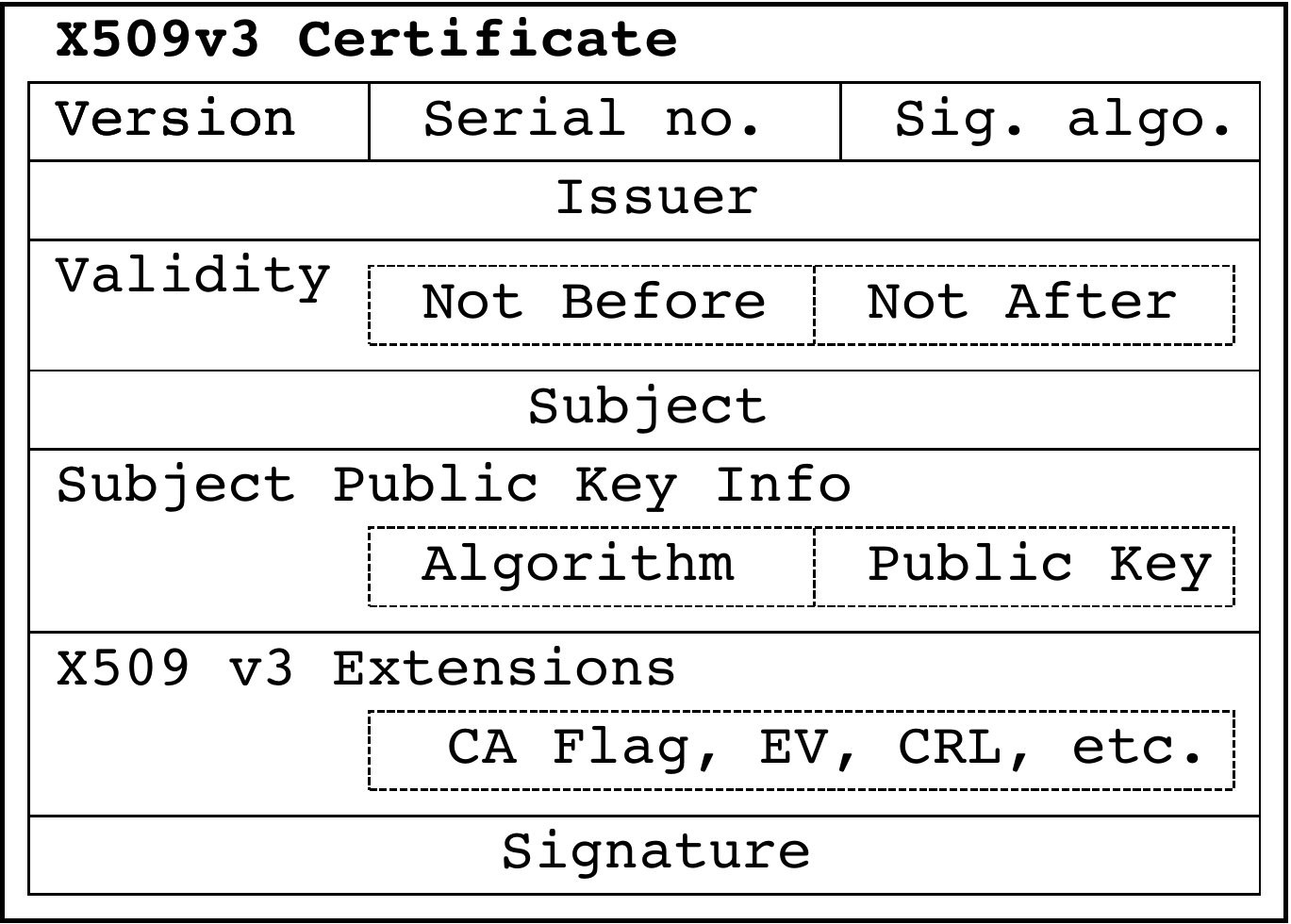}
  \caption{An X.509 certificate.}
  \label{fig:x509cert}
\end{figure}

\begin{figure}[tb]
  \centering
  \includegraphics[width=0.35\textwidth]{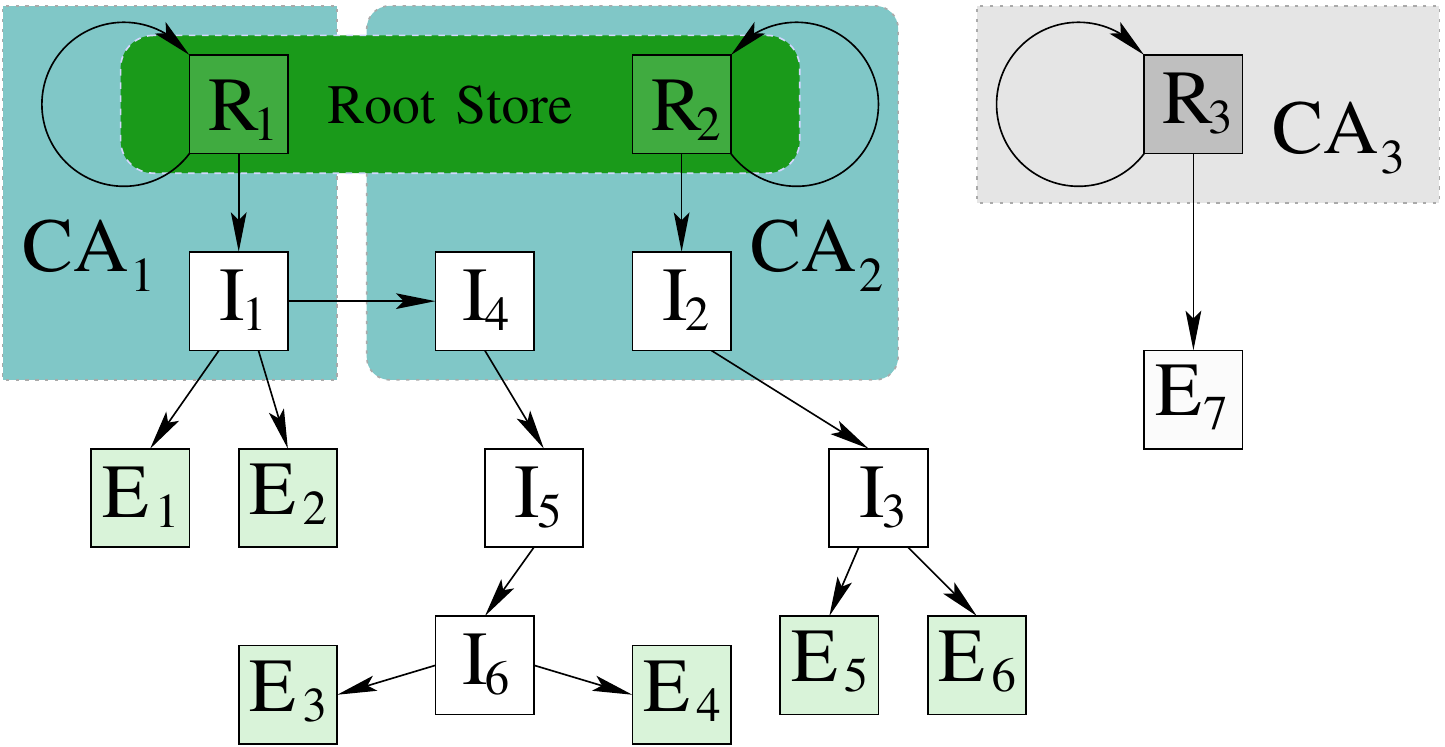}
  \caption{X.509 PKI showing the most relevant features: root certificates, intermediate certificates, end-host certificates, a root store and a certificate signed by an untrusted CA.}
  \label{fig:x509tree}
\end{figure}

\figurename~\ref{fig:x509cert} shows the format of an X.509 certificate, and
\figurename~\ref{fig:x509tree} shows a simplified PKI. The certificates of the
CAs form trust anchors, which are distributed with operating systems and user
agents like email clients (\eg Thunderbird) or web browsers.  For instance,
the Windows and OS X operating systems come with root stores supplied by the
vendors, who decide which CAs they include.  Software on this
OSes generally uses the OS-supplied trust anchors.  Mozilla takes a different
approach: their products have their own root store.

CAs can issue certificates directly (a practice that is thoroughly discouraged;
see Section \ref{sub:poorcapractice}) or via intermediate certificates. Trust
chains must not be broken, \ie missing intermediate certificates, chaining
to root certificates that are not in the root store, having expired
certificates in the chain \etc Self-signed certificates, where root and end-host certificates are the same, are a special case, which we discuss in Section
\ref{sub:correctchains}. Later in this paper, in Section
\ref{sec:securityanalysis}, we will discuss problematic PKI setups after going
through several observations from our measurements.

\subsection{Client-authentication methods}

The \ctos communication protocols examined in this paper generally authenticate
the initiator of the communication on the application layer and not in-band
during the \ssltls handshake. SMTP did originally not require authentication
for message submission (\ie user agent to mail server), but this was added
later to fight spam. Message transfer between MTAs (\ie transfer from the
source MTA to the destination MTA) does not require authentication of the
sender.

To choose the appropriate authentication mechanism, a client is supposed
to query the server for the mechanisms it supports (\eg using the EHLO command
with SMTP, or CAPABILITY for IMAP).  The server returns a list of supported
authentication mechanisms, sorted by preference, from which the client then
selects.

Some of the most widely used mechanisms, LOGIN and PLAIN~\cite{rfc4616},
transmit user credentials without further protection (independently of whether
there is an underlying \ssltls connection or not).  Some other mechanisms use
cryptographic functions to transmit a hashed version of the credentials (often
using deprecated hash functions such as MD5). An adversary who is able to
eavesdrop on the authentication process can potentially recover the
credentials.  Challenge-response mechanisms such as CRAM~\cite{rfc2195} and
SCRAM~\cite{rfc5802} (which also use HMAC) provide much better protection.
With these mechanisms, the password is never transmitted at all. In the case of
SCRAM, the password can even be stored in a salted format on server-side, and
hence not even a server compromise would reveal the true password to an attacker.

\subsection{Related work}
\label{sec:relatedwork}

A number of publications have studied the deployment of network security
protocols, with a focus on either the development of generic, large-scale
measurement methodologies or the measurement and analysis of the HTTPS and SSH
protocols.

Provos and Honeyman \cite{Provos2001} were probably the first to carry out
academic, large-scale scans of security protocols. Their work focused on SSH.
Later, Heidemann \etal carried out a census of Internet hosts
\cite{Heidemann2008}. Leonard and Loguinov \cite{Leonard2010} presented a
scanner capable of carrying out Internet-wide scans with proper randomisation
of target IP addresses. Durumeric \etal presented the fast \zmap scanner in
2013 \cite{Durumeric2013}. We used \zmap in our work.

Vratonjic \etal \cite{Vratonjic2011} carried out a scan of the top 1 million
hosts as determined by Alexa Inc. Holz \etal \cite{Holz2011} carried out scans
of the HTTPS ecosystem in a large-scale, long-term study over the duration of
18 months. The authors also used data from passive monitoring (using the Bro
Network Monitor). The study showed the poor state of the Web PKI and predicted
very little movement towards improvement.  More recently, Durumeric \etal
\cite{Durumeric2013a} presented an Internet-wide study of the HTTPS certificate
ecosystem; Huang \etal \cite{Huang2014} expanded on this in their investigation
of forward-secure cipher deployments in TLS.  Amann \etal \cite{Amann2013} and
Akhawe \etal \cite{Akhawe2013} carried out two studies that analysed the
aspects of trust relationships of the Web PKI and the occurrence and treatment
of error cases during certificate validation in popular implementations, again using
data from passive monitoring with Bro.

Some studies focused more on vulnerabilities in the wild. Heninger \etal
\cite{Heninger2012} studied data sets won with \zmap to investigate the cause
and distribution of weak RSA and DSA keys.  In their study of the Heartbleed
vulnerability, Durumeric \etal~\cite{Durumeric2014} also found email and XMPP
servers to be vulnerable.  Gasser \etal~\cite{Gasser2014} presented a
large-scale study of the deployment of SSH in 2014, with a focus on the
distribution of insecurely configured devices. 

There are not many studies that would focus on the use of \ssltls beyond HTTPS.
Concerning email, a recent
study~\cite{2015durumeric_empirical_analysis_mail_security} actively probed the
most popular email servers and observed the security of SMTP servers
interacting with Gmail over the duration of a year. The authors found that the
most popular providers did a decent job in setting up secure servers.  A paper
that was not yet published at the time of our initial submission also
investigated the security of email server setups \cite{Foster2015}. The authors
limited themselves to a relatively small number of servers. However, 
an important finding of theirs is that SMTP servers often do not verify the correctness
of a certificate in outgoing connections.  In our own study, we extend our
analysis to the whole Internet, but also to client-facing email retrieval
protocols and chat protocols. On a global scale, our findings are not as
reassuring as those for the most popular providers.

Finally, a number of online dashboards give some insight into the current
deployment of \ssltls: SSL Pulse for the most popular
websites\footnote{\texttt{https://www.trustworthyinternet.org/ssl-pulse/}}, Gmail
about their SMTP peers\footnote{\texttt{
https://www.google.com/transparencyreport/saferemail/}}, or the IM observatory
for XMPP servers\footnote{\texttt{https://xmpp.net/reports.php}}. The ICSI
Certificate Notary\footnote{\texttt{http://notary.icsi.berkeley.edu/}} also offers
an online, DNS-based query system that allows to check the validity of a given
X.509 certificate.

\section{Data collection}
\label{sec:datacollection}

We collected data using both active scans and passive measurement, \ie traffic
monitoring.  We use our scans to characterise global \tls deployment. The use
of passive monitoring data allows us to understand which specifics of \tls are
actually used; \eg which protocol versions and cipher suites are negotiated
between communication partners. Active scans are not as suitable for this purpose:
the responder chooses the cipher suite from the initiator's offers.

For email, we include all three \ssltls-variants of SMTP: SMTP with STARTTLS on
port 25, SMTPS on port 465, and SUBMISSION with STARTTLS on port 587. For IMAP
and POP3, we chose both the pure \ssltls as well as the STARTTLS variant. For
XMPP, we investigate both \ctos and \stos setups, in both STARTTLS and pure \ssltls
variants.  For IRC, we only investigate the \ctos communication\footnote{Recall
that \stos communications are not standardised.}. We limit our IRCS scan to the
most common port, 6697, and probe for IRC STARTTLS support on the default IRC
port, 6667.

\subsection{Active scans} \setcounter{paragraph}{0}

\begin{table*}[bt]

  \caption{Description of our active scan dataset containing hosts
    listening on ports, successful handshakes, end-host and intermediate
    certificates. Entries marked with $\dag$ used \starttls, and those with
  $\ddag$ allowed fallback to \ssl. S2S is short for \stos, C2S for \ctos.}

  \label{table:hosts}

  \centering
  \small

  \begin{tabular}{lrrrrrr}
    \toprule
    \multicolumn{1}{c}{\textbf{Protocol}}	& \multicolumn{1}{c}{\textbf{Port}}	& \multicolumn{1}{c}{\textbf{Period}}	& \multicolumn{1}{c}{\textbf{No. hosts}}	& \multicolumn{1}{c}{\textbf{Successful \ssltls}}	& \multicolumn{1}{c}{\textbf{Unique end-host-certs}} 	& \multicolumn{1}{c}{\textbf{Intermediate certs (unique)}}	\\
	\midrule                                                                        
	SMTP$^{\dag,\ddag}$			& 25					& 7/27--7/28					& \num{12488000}			& \num{3848843} (30.82\%)			& \num{1373751} (35.69\%)		 		& \num{2243846} (\num{23462}, 1.05\%)			\\
	SMTPS$^{\ddag}$				& 465					& 7/22--7/23					& \num{7234817}				& \num{3437382} (47.51\%)			& \num{800574} (23.29\%)				& \num{2583786} (\num{10357}, 0.4\%)				\\
	SUBMISSION$^{\dag,\ddag}$			& 587					& 7/27					& \num{7849434}				& \num{3378009} (43.03\%)			& \num{753691} (22.31\%)				& \num{2580305} (\num{16070}, 0.62\%)				\\
	IMAP$^{\dag,\ddag}$			& 143					& 7/25--7/26					& \num{8006617}				& \num{4076809} (50.91\%)			& \num{1024757} (25.14\%)				& \num{2406987} (\num{12913}, 0.54\%)				\\
	IMAPS					& 993					& 7/09--7/11					& \num{6297805}				& \num{4121108} (65.43\%)			& \num{1053110} (25.55\%)				& \num{2791451} (\num{16700}, 0.6\%)				\\
	POP3$^{\dag,\ddag}$			& 110					& 7/26						& \num{8930688}				& \num{4074211} (45.62\%)			& \num{998013} (24.5\%)					& \num{2325032} (\num{10135}, 0.44\%)			\\
	POP3S					& 995					& 7/10--7/12					& \num{5186724}				& \num{2797300} (53.93\%)			& \num{747508} (26.72\%)				& \num{1795814} (\num{7876}, 0.44\%)				\\
	\midrule                                                                        
	IRC$^{\dag}$				& 6667					& 8/02--8/04					& \num{2573207}				& \num{3709} (0.14\%)				& \num{3003} (80.97\%)					& \num{638} (\num{84}, 13.17\%)					\\
	IRCS					& 6697					& 7/17--7/18					& \num{1948656}				& \num{8661} (0.44\%)				& \num{6332} (73.11\%)					& \num{2551} (\num{315}, 12.35\%)				\\
	XMPP, C2S$^{\dag,\ddag}$		& 5222					& 7/29--7/30					& \num{2188813}				& \num{53544} (2.44\%)				& \num{38916} (63.61\%)				 	& \num{5927} (\num{1913}, 32.28\%)				\\
	XMPPS, C2S				& 5223					& 7/13--7/14					& \num{2223994}				& \num{70441} (3.16\%)				& \num{38916} (55.25\%)				 	& \num{32629} (\num{2773}, 8.5\%)				\\
	XMPP, S2S$^{\dag,\ddag}$		& 5269					& 7/31--8/01					& \num{2459666}				& \num{9780} (0.39\%)				& \num{6221} (63.61\%)					& \num{5927} (\num{1913}, 32.28\%)				\\
	XMPPS, S2S$^{\ddag}$			& 5270					& 7/24						& \num{2046204}				& \num{1693} (0.08\%)				& \num{1146} (67.69\%)					& \num{783} (\num{147}, 18.77\%)				\\
	\midrule                                                                        
	HTTPS                          		& 443     				& 6/30--7/09    				& \num{42676912}    			& \num{27252853} (63.85\%)   			& \num{8598188} (31.55\%)  				& \num{24555475} (\num{227321}, 0.93\%)				\\ 
    \bottomrule
  \end{tabular}
\end{table*}

In this section, we describe the process we used to perform our active scans.
We also explain some insights we gained and some peculiar phenomena we
encountered when scanning.

\paragraph{Scanner}

Our scanner consists of two parts. The first is the
\texttt{zmap}~\cite{Durumeric2013} network scanner, which we used to determine
IP addresses that had ports of interest open. We scanned the entire routable
IPv4 space\footnote{Appropriate ways to scan IPv6 are an open research topic.},
using a BGP dump from the Oregon collector of
Routeviews\footnote{\texttt{http://www.routeviews.org}} as a whitelist of routable
prefixes. We ran our scanning campaigns over several weeks, from 2015-06-09
through 2015-08-04. Due to time-sharing constraints on the scanning machine, we
had to run the scans at different speeds, resulting in scans of different
durations, as summarised in Table~\ref{table:hosts}. In general, scans lasted
roughly 20-36 hours. We refrained from scanning at line speed
(although this is possible with our setup) to reduce our scans' impact.

The second part of our scanner is a component that starts an array of
OpenSSL client instances, collects their output, and stores it in a database. We patched
the STARTTLS implementations of OpenSSL as the current version does not
follow the RFCs. More specifically, the current OpenSSL client does not query the server
capabilities and ignores a server's refusal to negotiate \ssltls. Furthermore,
OpenSSL did not yet support \starttls for IRC, either.

We used a blacklist of IP ranges generated during past
scans~\cite{Gasser2014,Schlamp2015}.  At the time of writing, it contains 177
entries covering 2.6 million addresses (about 0.08\% of the routable space).
Entries were computed from both automated and personal emails that reached us
and complained about the scans.

\paragraph{Scanned protocols}

Table~\ref{table:hosts} gives an overview of our dataset from active scans. It
shows the number of hosts responding to connection attempts as well as the
number of hosts to which a successful \ssltls connection could be established.
The table also lists the number of unique end-host certificates that we
encountered on all machines in the respective scans. Furthermore, it contains
the number of total and unique intermediate certificates encountered in the
scans. Note that many servers seem to have a \ssltls port open, yet do not
carry out successful \ssltls handshakes. This phenomenon has been observed
before for HTTPS \cite{Holz2011,Durumeric2013}; we encounter it again for email
and chat.

Previous scans performed by us show that servers that support only \ssl are
very rare today. Modern Debian-based systems do not even include it in the
default OpenSSL binary they ship. Initially, we followed their lead and did not
try to connect with an optional fall-back to \ssl.  However, we revised that
decision after inspecting data from the passive monitoring and deciding we
wanted to allow for some comparisons. We thus enabled fall-back to \ssl for the
remainder of our scans.

\paragraph{Background noise} We observed a phenomenon which has also been
mentioned before by the \zmap community: independently of the port one
chooses to scan Internet-wide, there is always a number of hosts that reply to
SYN packets without carrying out a full TCP handshake later.  We verified this
by scanning five arbitrarily chosen ports (1337, 7583, 46721, 58976, 65322) and
sending out 100M probes each time. We scanned twice with different seeds for
each port.  Every time, the average response rate was 0.07--0.1\%. When
scanning protocols with very low deployment, it is important to keep this
phenomenon in mind as one of the causes for failed \ssltls handshakes.  This is
particularly important to consider for less-used protocols such as IRC or XMPP.

\subsection{Passive collection}
\setcounter{paragraph}{0}

\begin{table}[bt]
  \caption{Connections and servers in passive scans. Entries~marked~with~$\dag$ used \starttls. S2S is short for \stos, C2S for \ctos.}
  \label{tab:passivenums}
  \begin{center}
    \begin{tabular}{lrrr}
    \toprule
    \multicolumn{1}{c}{\textbf{Protocol}}	& \multicolumn{1}{c}{\textbf{Port}}	& \multicolumn{1}{c}{\textbf{Connections}}	& \multicolumn{1}{c}{\textbf{Servers}}	\\
    \midrule
    SMTP$^{\dag}$				& 25					& \num{3870542}					& \num{8626}				\\
    SMTPS					& 465					& \num{37306}					& \num{266}				\\
    SUBMISSION$^{\dag}$			& 587					& \num{7849434}					& \num{373}				\\
    IMAP$^{\dag}$				& 143					& \num{25900}					& \num{239}				\\
    IMAPS					& 993					& \num{4620043}					& \num{1196}				\\
    POP3$^{\dag}$				& 110					& \num{18774}					& \num{110}				\\
    POP3S					& 995					& \num{159702}					& \num{341}				\\
    \midrule
    IRC$^{\dag}$				& 6667					& \num{53}					& \num{2}				\\
    IRCS					& 6697					& \num{18238}					& \num{15}				\\
    XMPP, C2S$^{\dag}$				& 5222					& \num{13517}					& \num{229}				\\
    XMPPS, C2S					& 5223					& \num{911411}					& \num{2163}				\\
    XMPP, S2S$^{\dag}$				& 5269					& \num{175}					& \num{2}				\\
    XMPPS, S2S					& 5270					& \num{0}					& \num{0}				\\
    \bottomrule
    \end{tabular}
  \end{center}
\end{table}

For our passive measurements, we examined nine days of traffic of the Internet
uplink of the University of California at Berkeley, which has a 10\,GE uplink
with a peak traffic of more than 7\,GB/s each way.

\paragraph{Traffic monitoring and capture}

We used the Bro Network Security
Monitor\footnote{\texttt{http://www.bro.org}}~\cite{paxson99bro} to gather
information about all outgoing \ssltls sessions.  In a default installation, Bro
already offers deep visibility into standard \ssltls traffic, extracting
certificates and meta-information like cipher and key use.  For this work, we
extended Bro to also work with protocols using \starttls.
We added support for STARTTLS for the SMTP, POP3, IRC, XMPP, and IMAP
protocols.

We also use Bro to extract the server's offered authentication capabilities for
all outgoing SMTP, POP3, and IMAP sessions, which allows us to deduce how many
of the contacted servers support \starttls. We added support
for capabilities to the IMAP protocol analyser we created for this work;
support was already present in Bro for SMTP and POP3 capabilities.

Our passive dataset was collected from 2015-07-29 to \mbox{2015-08-06}. We observed a
total of 9,730,095 \ssltls connections on the monitored ports. The connections
were established to 12,637 unique destination IP addresses with 10,294 distinct
Server Name Indication (SNI) values and 10,164 unique end-host certificates.
Table~\ref{tab:passivenums} shows the number of connections and servers
encountered per port.

Please note that our passive data set exhibits artefacts of the collection
process that are beyond our control. As our data is collected at the Internet
uplink of one university, it is potentially biased. We assume that, due to the
high number of students with diverse cultural backgrounds, the traffic we see
is similar to traffic in other parts of the world, however.

\paragraph{Ethical considerations}

We are aware of the ethical considerations that must be taken into account when
observing passive traffic. This research strives to understand the interplay
between server and client software at the technical level and does not concern
any human subjects. For the \ssltls measurements, the information that we save
is constrained to information in the \ssltls handshake without analysing any
later connection payload data. The campus administration has approved this data
collection. For the capability measurements, only automatic server capability
replies were recorded, which do not contain any personally identifiable
information. In addition, the University IRB takes the position that
IP~addresses, which were also recorded for this measurement, are not treated as
personally identifiable.

\paragraph{Unusual traffic on standard ports}

While analysing the \tls extensions sent by clients, we noticed that there are
4,584 connections that send the Application-Layer Protocol Negotiation (ALPN)
extension, which is used to negotiate protocols like HTTP2 and SPDY. Closer
examination shows that 2,703 of these connections going to six servers indeed contain
values that point to them being HTTPS servers, running on port 993 as well as
110. Manually connecting to a few of these IP addresses shows that they are
Squid proxy servers running on non-standard ports. The remaining 1,881
connections to 780 hosts all have a destination port of 5223. The ALPN in these
cases indicates a value of \texttt{apns-security-v1} and
\texttt{apns-security-v2}, terminating at nodes for the Apple push notification
service. We are not sure what software causes these connections.  Further
traffic analysis also reveals that our data set contains 3,728 certificates,
from 9,082 connections to 110 servers, indicating that they are used by the Tor
service. We excluded all these servers from further analyses.

\paragraph{\ocsp stapling}

Another interesting finding is the adoption of \ocsp~\cite{rfc4806} stapling by
email servers.  \ocsp stapling allows \tls servers to send a proof that their
certificate is currently still valid and has not been revoked. This is part of
the \tls handshake if the client signals support for the extension.  We
encountered 836 connections using \ocsp stapling, terminating at 64 different
servers. The majority of these (706 connections and 58 servers) were on port
993 (IMAPS).

\section{Security analysis}
\label{sec:securityanalysis}

We now analyse our datasets from a security perspective. Specifically, we
examine how appropriately servers are configured. We look at basic parameters
such as the ciphers in use and also consider PKI-related specifics, such as
whether the offered certificates are valid and linkable to CAs present in the
Mozilla root store, the amount of key and certificate reuse, and whether CAs
follow best practices in issuing certificates. We finally study authentication
methods offered to clients. Where applicable, we compare our findings for email
and chat protocols with results from our HTTPS scan.

\subsection{Use of STARTTLS vs. direct \ssltls}
\label{sec:starttls}

\begin{table}[t]
  \addtolength{\tabcolsep}{-4pt}
  \caption{STARTTLS support and use. Passive monitoring allows to differentiate server-side support from client--server connections which were actually negotiated. S2S is short for \stos, C2S for \ctos.}
  \label{table:upgradedconnections}
  \begin{center}
    \begin{tabular}{lrrrr}
      \toprule
      						& \multicolumn{1}{c}{\textbf{Active probing}}		& \multicolumn{3}{c}{\textbf{Passive monitoring}}	\\
      \cmidrule{2-5}
      						& \multicolumn{1}{c}{\textbf{Supported}}		& \multicolumn{1}{c}{\textbf{Supporting}}	& \multicolumn{1}{c}{\textbf{Offering}}	& \multicolumn{1}{c}{\textbf{Upgraded}}	\\
      \multicolumn{1}{c}{\textbf{Protocol}}	& \multicolumn{1}{c}{\textbf{\& upgraded}}		& \multicolumn{1}{c}{\textbf{servers}}	& \multicolumn{1}{c}{\textbf{connections}}	& \multicolumn{1}{c}{\textbf{connections}}	\\
      \midrule
      SMTP					& 30.82\%						& 59\%							& 97\%							& 94\%			\\
      SUBMISSION				& 43.03\%						& 98\%							& 99.9\%						& 97\%			\\
      IMAP					& 50.91\%						& 77\%							& 70\%							& 44\%			\\
      POP3					& 45.62\%						& 55\%							& 73\%							& 62\%			\\
      \midrule
      IRC					& 0.14\%						& --							& --							& --			\\
      XMPP, C2S					& 2.44\%						& --							& --							& --			\\
      XMPP, S2S					& 0.39\%						& --							& --							& --			\\
      \bottomrule
    \end{tabular}
  \end{center}
\end{table}

As mentioned in Section~\ref{sec:background}, email and chat protocols can be
secured with \ssltls either by using \ssltls on a dedicated port or by
upgrading a TCP connection via STARTTLS.

Table~\ref{table:hosts} shows how many hosts supported \ssltls directly. We
also measured support for STARTTLS in our active and passive scans (see
table~\ref{table:upgradedconnections}). Our data shows that, depending on the
application-layer protocol, about 30 to 51\% servers offer \starttls. The
STARTTLS extension is also often used in practice. While popular servers seem
to support the extension (and thus most connections contain an offer to use
it), the results for SMTP, IMAP, and POP3 do show that there is a significant
number of servers without support. At least in the case of IMAP and POP3, one
can also see that, in a considerable number of cases, connections are not
upgraded although the server would support it.

\subsection{\ssltls versions---deployment and use}
\label{sub:ssl3use}

\begin{table}[t]
  \caption{Negotiated protocol versions from active scans with \ssl activated and passive monitoring.}
  \label{tab:protocolversions}
  \begin{center}
    \begin{tabular}{lrr}
      \toprule
						& \multicolumn{1}{c}{\textbf{Active probing}}		& \multicolumn{1}{c}{\textbf{Passive monitoring}}	\\
      \multicolumn{1}{c}{\textbf{Version}}	& \multicolumn{1}{c}{\textbf{Negotiated with server}}	& \multicolumn{1}{c}{\textbf{Observed connections}}	\\
      \midrule
      \ssl					& 0.02\%						& 1.74\%						\\
      \tls~1.0					& 39.26\%						& 58.79\%						\\
      \tls~1.1					& 0.23\%						& 0.1\%							\\
      \tls~1.2					& 60.48\%						& 39.37\%						\\
      \bottomrule
    \end{tabular}
  \end{center}
\end{table}

Ideally, only the latest version of \tls (1.2) should be used. Previous
versions, especially \ssl, have vulnerabilities, many of which are listed in
RFC~7457~\cite{rfc7457}.

\tablename~\ref{tab:protocolversions} shows how often the different \ssltls
protocol versions were chosen by servers in active scans (in scans that
supported \ssl). Note that we did not scan for \oldssl.
\tablename~\ref{tab:protocolversions} also shows protocol versions observed in
use in our passive monitoring dataset. No connections using \oldssl were
encountered.

Our result shows that just 0.03\% of scanned servers only support the old and
relatively insecure \ssl---all others preferred one of the stronger \tls
versions. However, the percentage of connections actually using \ssl in our
passive dataset is much higher (1.74\%). There are two possible reasons for
this. Either clients connect preferentially to less secure servers---this would
not be in line with our results for STARTTLS support in the previous section.
Or there is a significant number of clients that offer \ssl only, \eg because
they are outdated.

\subsection{Cipher use}
\label{sec:ciphers}

\begin{figure}[t!]
  \centering
  \includegraphics[width=0.5\textwidth]{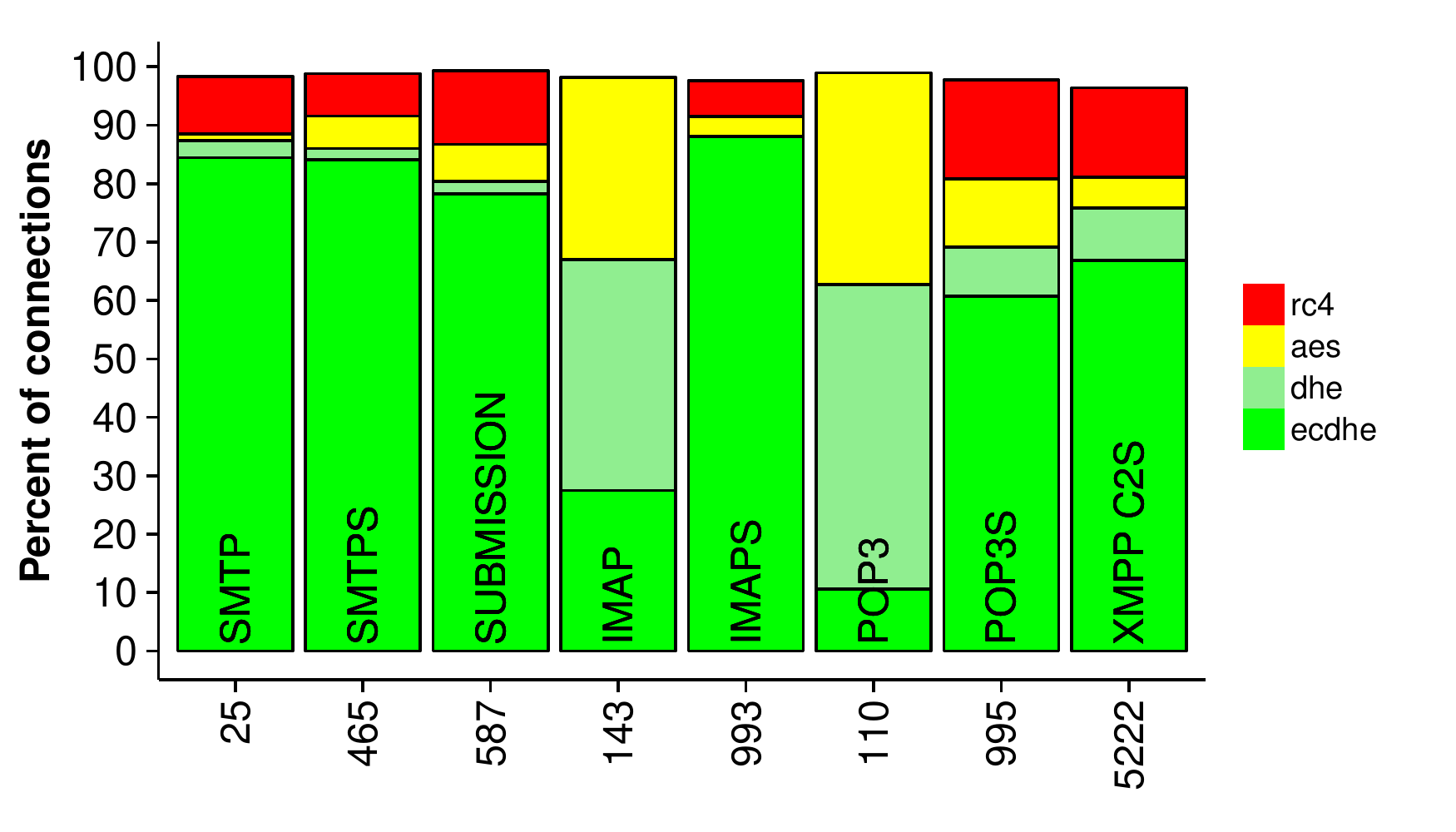}
  \caption{Use of PFS ciphers by port. Red and yellow indicate that PFS is not used.}
  \label{fig:server-ciphers}
\end{figure}

In \ssltls, the server chooses the symmetric cipher to use, based on a list of
ciphers that the client suggests. Determining which ciphers a server supports
would require many connections to test all ciphers individually.  Given that
many of those suites may never be negotiated, this is a poor trade-off in
terms of good Internet citizenship versus lessons that can be learned.

We thus use passively monitored data to investigate which ciphers are actually
negotiated in practice. Due to the high number of different cipher suites
occurring in the wild---35 in our dataset---we group the ciphers into
categories that show their relative strengths.  Figure~\ref{fig:server-ciphers}
shows the different categories.  The categories ECDHE and DHE contain suites
that use forward-secure (PFS) ciphers, either using elliptic curve or
modular Diffie--Hellmann key exchanges.  The categories AES and RC4 contain
connections without PFS support that use either the AES or RC4 cipher. Other
categories with a use of less than 1\% of connections were omitted (an example
for this are connections using the Camellia cipher). Connections on ports 6679
and 6667 overwhelmingly use ECDHE ciphers, and those on port 5269
overwhelmingly use DHE ciphers.  These ports were excluded from the figure for
brevity.  Figure~\ref{fig:server-ciphers} shows that there is still a
surprisingly large amount of connections on some ports that use the RC4 stream
cipher.

Looking at the elliptic curves that are used in ECDHE key exchanges reveals
that 97.2\% of connections use the \texttt{secp256r1} curve, followed by 2\%
using \texttt{secp384r1} and 0.78\% using \texttt{sect571r1}. All of these
curves are considered to be at least as strong as 2048 bit RSA, raising no
immediate security concerns. This result is similar to earlier results
concerning server support for different curves \cite{Huang2014}.

Examining the Diffie--Hellmann parameter sizes for the DHE connections reveals
that 76\% of the connections use a parameter size of 1024 bit, 22\% of 2048 bit,
and 1.4\% of 768 bit. While this is an improvement in comparison to earlier
studies, which measured more than 99\% of hosts only supporting 1024 bit keys
and below (see~\cite{Huang2014}), this is still relatively poor as 
parameter sizes below 2048 bits are discouraged today.


\subsection{Certificate chain validity}
\label{sub:correctchains}
\setcounter{paragraph}{0}

\begin{figure}[tb]
  \centering
  \includegraphics[width=0.5\textwidth]{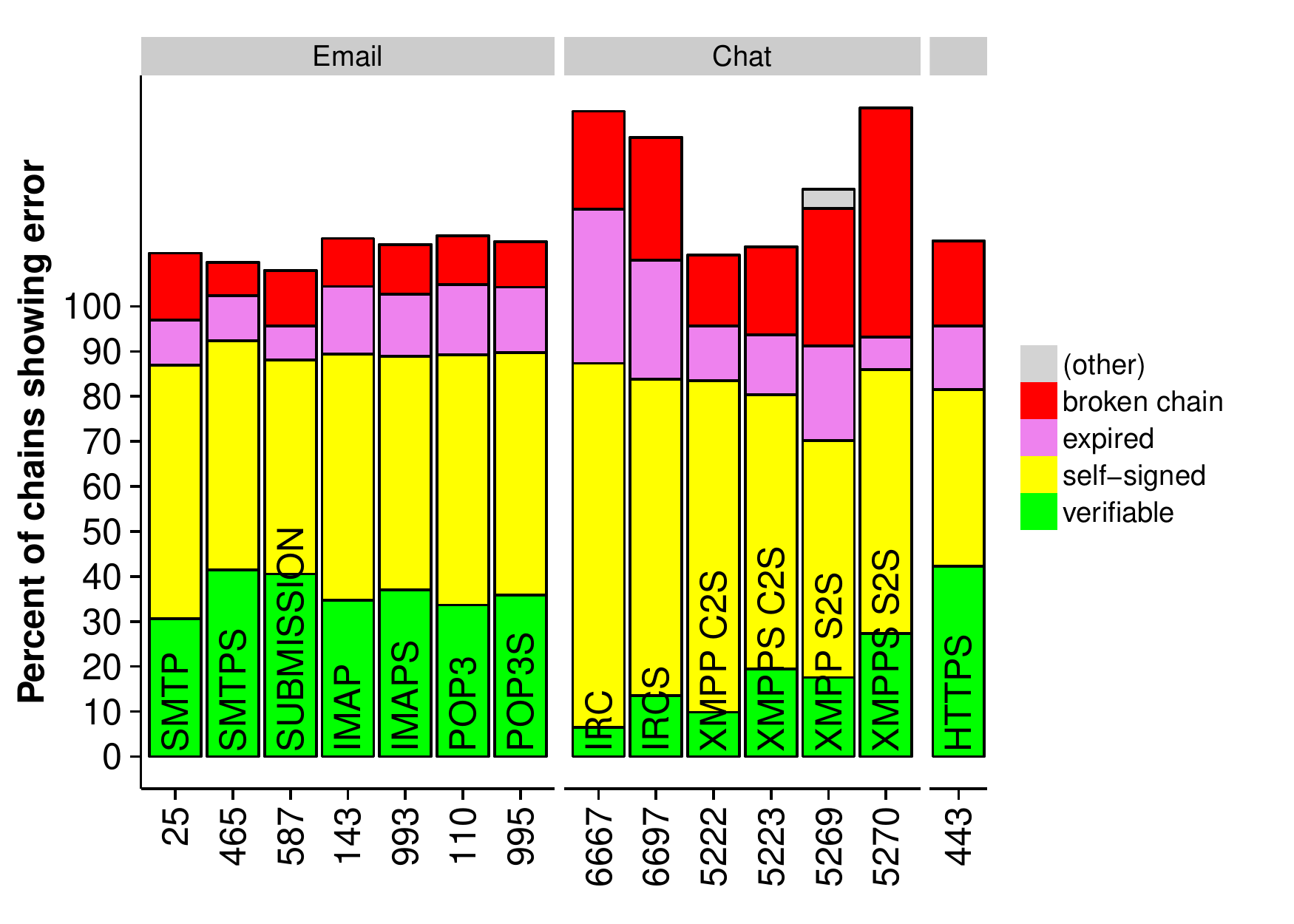}
  \caption{Common errors in certificate chains, active scans. Note that chains may exhibit more than one error, which we capture in this figure. Thus, the results may add up to more than 100\%.}
  \label{fig:chainvalidationactive}
\end{figure}

\ssltls servers send a certificate chain in the handshake that consists of the
host's certificate and potentially intermediate and CA certificates. It is
common to omit the CA certificate as it already has to be part of the local
root store.
Chains can exhibit several types of errors---certificates may be expired, host
certificates may not chain up to a root certificate in the root store,
intermediate certificates can be missing, \etc
A particularly common case are self-signed certificates, where issuer and
subject of the certificate are the same.  While technically not an error, these
certificates can only serve the use case where a private server operator does
either not care about authenticated encryption (and thus often uses some
standard certificate as supplied in, \eg Linux distributions) or issues a
certificate to herself and configures her clients to accept it.\footnote{Many
    clients allow to do this by storing an exception for the host and
certificate on the first connection, thus making all subsequent connections
secure as the stored certificate is compared against the one the server sends.}

\paragraph{Deployed vs. used services}

\begin{figure}[tb] \centering
\includegraphics[width=0.5\textwidth]{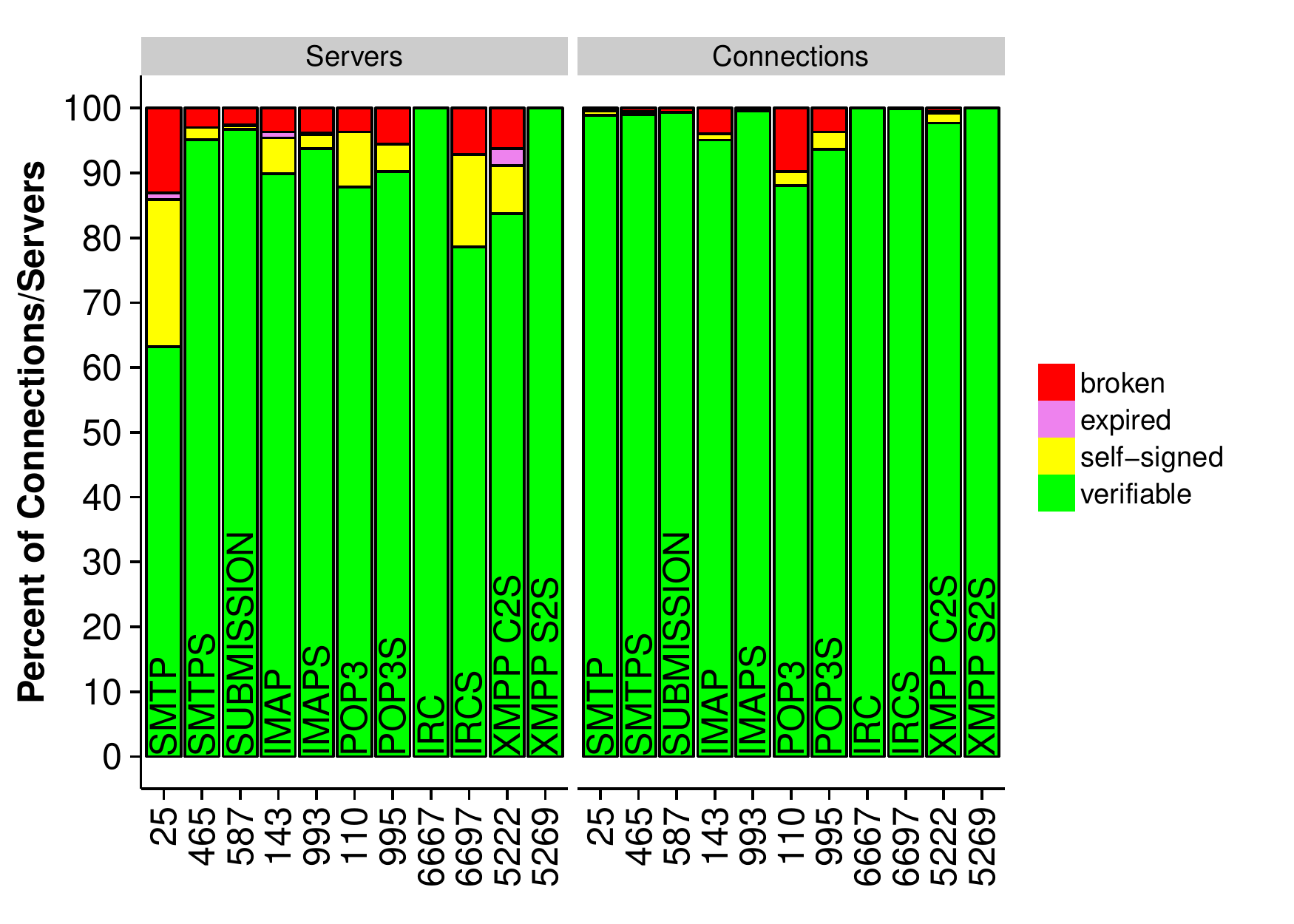} \caption{Common
errors in certificate chains, passive scans. Only the primary error (as reported by OpenSSL) is shown.}
\label{fig:chainvalidationpassive} \end{figure}

We show the most common certificate errors we encountered in our active scans in
\figurename~\ref{fig:chainvalidationactive}.
\figurename~\ref{fig:chainvalidationpassive} shows validation results for our
passive monitoring run by servers (\ie counting every server once) and
weighted by connections (\ie counting each server weighted by the amount of
connections that we saw).

This data set contains 295 cases where the same IP and port serves more than
one certificate chain. Examples for this are Google and other company mail
servers, servers where only CA certificates were updated while the end-host
certificate was left unchanged, and servers renewing their end-host
certificates. We examined the certificates sent by these servers and found that
they all share the same validity characteristics (\ie in all cases either all
of the certificates sent by a host were valid or invalid).

\figurename~\ref{fig:chainvalidationactive} shows that the ratio of verifiable
chains is between 30-40\% across all email protocols. This is much lower than
what has been reported for web sites on the Alexa Top 1 million list (around
60\%)~\cite{Holz2011}, but much more in line with what has been found for the
Web PKI as a whole~\cite{Durumeric2013a}.  For comparison, we also included the
values for HTTPS. Looking at the data from passive monitoring, we note that the
number of correctly validating chains is much higher when only considering
servers that actually receive connections, and even more so when weighting this
with the number of connections, as shown in
\figurename~\ref{fig:chainvalidationpassive}. This suggests that the operators
of the most popular services do a substantially better job at properly
configuring their server for use with \ssltls.

\paragraph{Invalid certificates}

Self-signed certificates are the major source of non-verifiable certificate
chains in our measurements. As mentioned above, clients that wish to
authenticate servers configured with such certificates must have
out-of-band knowledge about the correctness of the certificate. Note
that this approach only works where a self-signed certificate was created by the
administrator---default certificates, as they are often shipped with software
bundles, are useless for authentication as a copy of the private key is also 
shipped with the bundle. This is one case for certificate reuse, which we
discuss below.

Certificate chains can be broken in a number of ways---\eg missing intermediate
certificates, using CA certificates that are not in the root store, \etc We
grouped these errors together and found that their number is relatively low at
10-15\%.  Our result shows that, just as in the Web PKI, there are many
mistakes that can be made in certificate deployment. The number of expired
certificates, which we consider separately, is well within previously reported
ranges~\cite{Durumeric2013a, Holz2011}, showing that there is little difference
between email and web protocols in this regard.  We also found some further
errors in certificate chains that we classified as `other'---these are rare and
sometimes somewhat arcane\footnote{A full list of possible errors can be found
on the OpenSSL homepage; \texttt{https://www.openssl.org/docs/apps/verify.html}.}.
Just as in previous scans~\cite{Holz2011}, we found only a single-digit number
of cases with broken signatures.

Looking at the different protocols in
\figurename~\ref{fig:chainvalidationactive}, we see a difference between the
email protocols and the chat protocols. While SMTPS and SUBMISSION have the
highest (yet still unsatisfactory) percentage of verifiable certificate chains
(and IMAP, POP3, and SMTP are trailing not too far behind), the numbers are
much lower for XMPP and especially IRC. SMTP also has a much lower rate of
verifiable certificate chains in our passive scans, at least when not weighting
by number of connections: an indication that message protection in a number of
\stos communications is likely to be at higher risk (although once again,
popular servers seem to be properly configured).  This is a serious problem, which is also compounded by
the findings of a recent study that ran in parallel to ours \cite{Foster2015}:
the authors found that the servers in their study did not verify certificates
in outgoing connections at all. It is thus reasonable to assume that many SMTP
\stos connections are not secure.

A staggering number of IRC servers
seems to use self-signed certificates, or deploy broken or expired chains. This
puts private (person-to-person) IRC messaging as well as password transfers at
risk. We study the case of XMPP separately below.

\paragraph{The case of XMPP}

The vast majority of certificates deployed for the XMPP \ctos services 
(5222 and 5223) are self-signed. However, an inspection of typical common names
for these certificates shows that the corresponding servers are most likely
parts of proprietary deployments and not intended for general use. The
corresponding subjects for XMPP on port 5222 are shown in
\tablename~\ref{tab:cert-occ-cn-all}. For XMPPS on port 5223, 48\% were from a
Content Distribution Network (\texttt{incapsula.com}), 12\% from Apple's push
service (\texttt{courier.push.apple.com}) and another 8\% by a Samsung push
service (\texttt{*.push.samsungosp.com}). The remaining certificates have
shares between 2 and 5\% and contain variations of the subjects
\texttt{hub.clickmyheart.net}, \texttt{icewarp.com} and \texttt{ejabberd}---a
popular XMPP implementation. We thus conclude that this port is often used for
push services, rather than instant messaging.

Consulting our passive data set confirms this conclusion. 90\% (826,822) of
port 5223 connections to 1,282 servers use a SNI containing
\texttt{push.apple.com}, with all but two of the server IP addresses residing
in Apple's IP space\footnote{The remaining two addresses with one connection
    per address use an IPv6 target address in an address space a network
provider uses for NAT64; we assume these addresses also get redirected to Apple
servers in some way.}. 73,465 more connections target the Samsung service
mentioned above, pushing the connection numbers to these services beyond 99\%
of all port 5223 connections. Our passive observations also show that the
majority of \ctos connections have verifiable chains. This is also true when
looking at the distribution for servers only, albeit to a lesser degree. Once again we see a
preferential use of servers with better-than-common security.

For the \stos ports, which are used to relay XMPP messages, we found broken
(and not self-signed chains) to be slightly more common in our active scans 
(but notably not in our passive data). It
is difficult to arrive at a strong conclusion here.  The slightly lower
percentage of self-signed certificates may hint at conscious certification
choices made for \stos communication. Since XMPP is also used in proprietary
products (not meant for public access), operators may have chosen to use
private CAs instead of acquiring certificates from commercial CAs. If true,
we did not capture such communication in our passive observations.

\subsection{Key and certificate reuse} \setcounter{paragraph}{0}

\paragraph{Certificate reuse}

\begin{figure}[tb]
  \centering

  \subfloat[All certificates]{
    \includegraphics[width=0.5\textwidth]{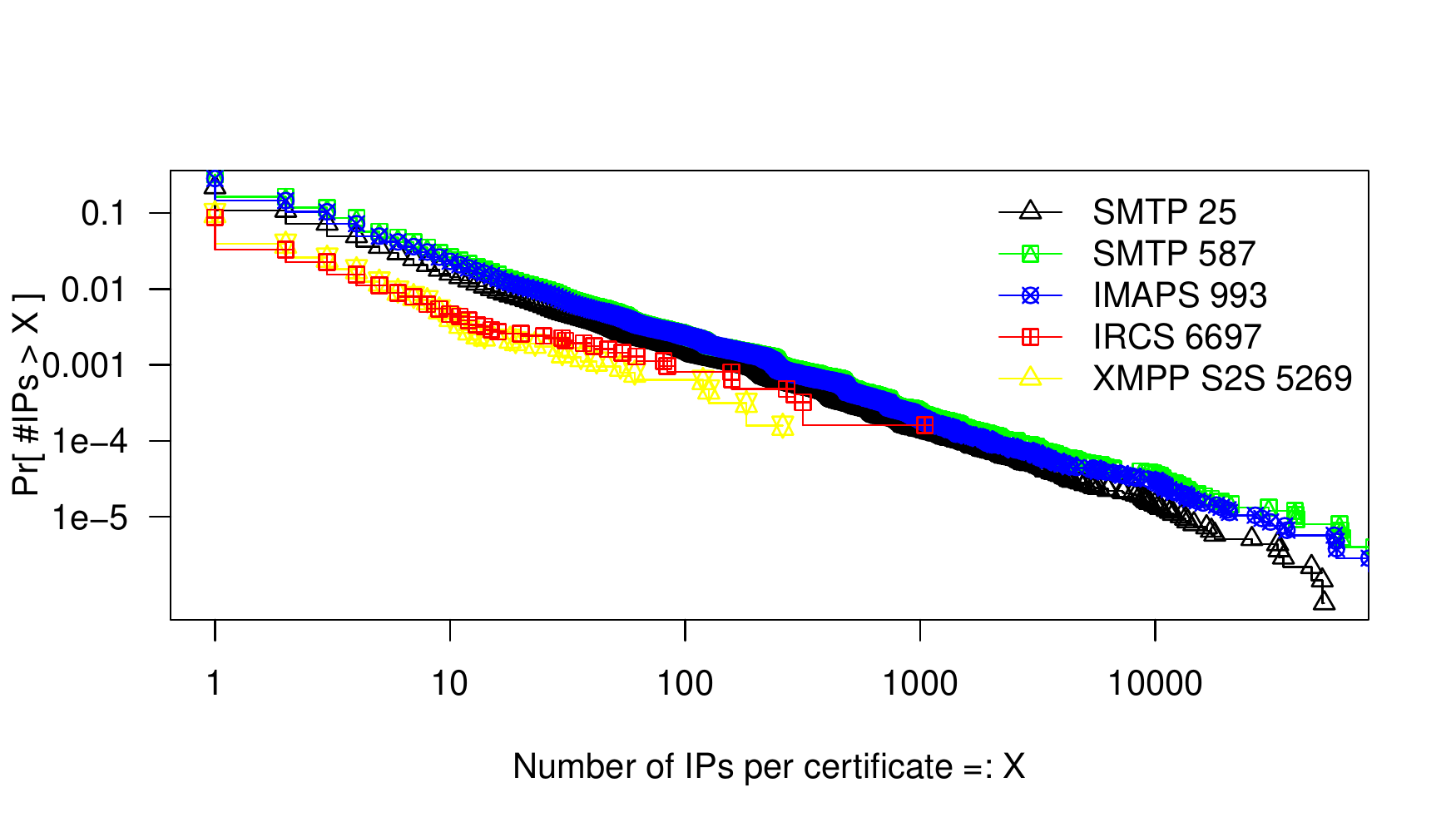}
    \label{fig:cert-occ-all}
  }

  \subfloat[Valid certificates only]{
    \includegraphics[width=0.5\textwidth]{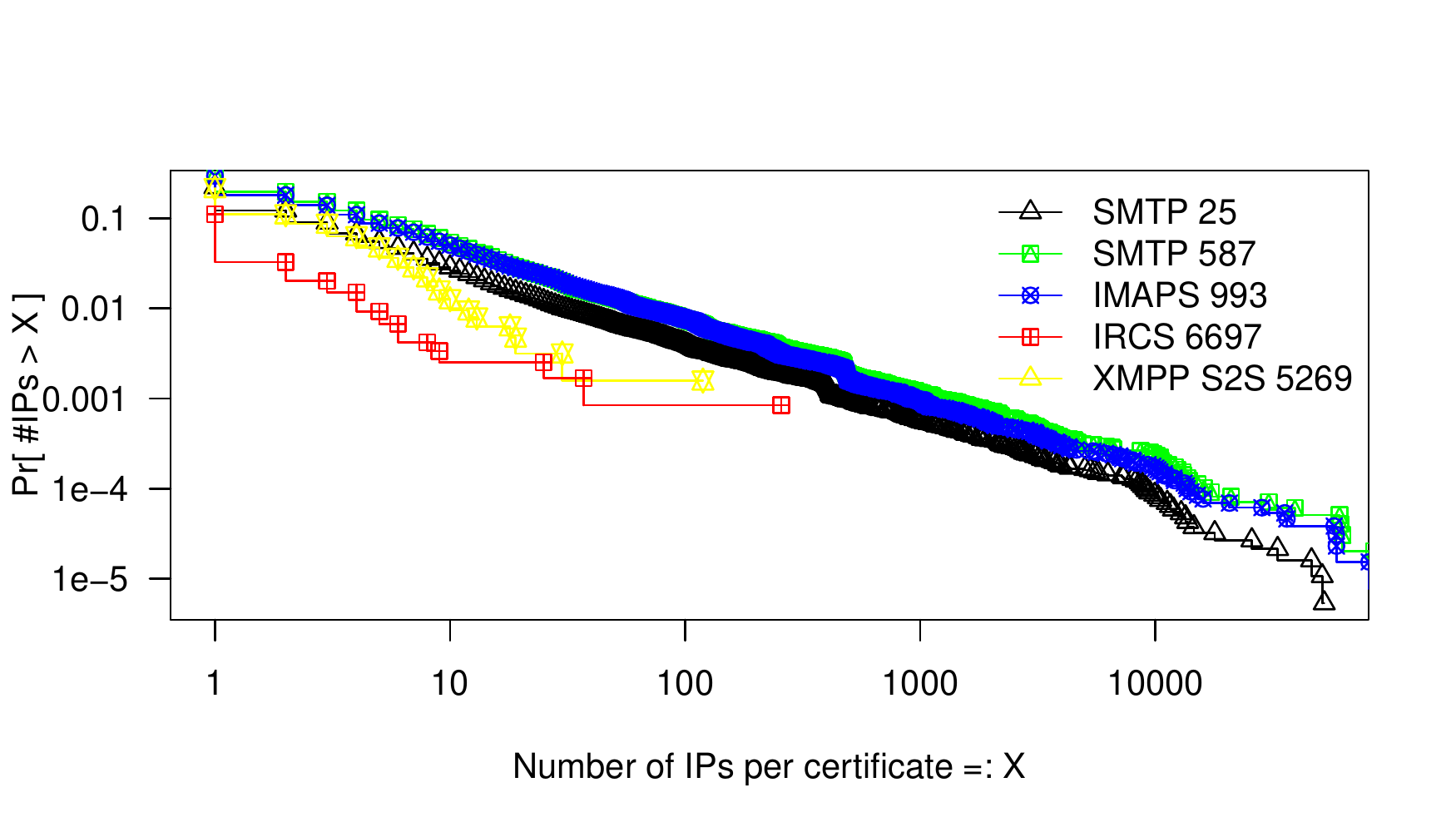}
    \label{fig:cert-occ-valid}
  }

  \subfloat[Self-signed certificates only]{
    \includegraphics[width=0.5\textwidth]{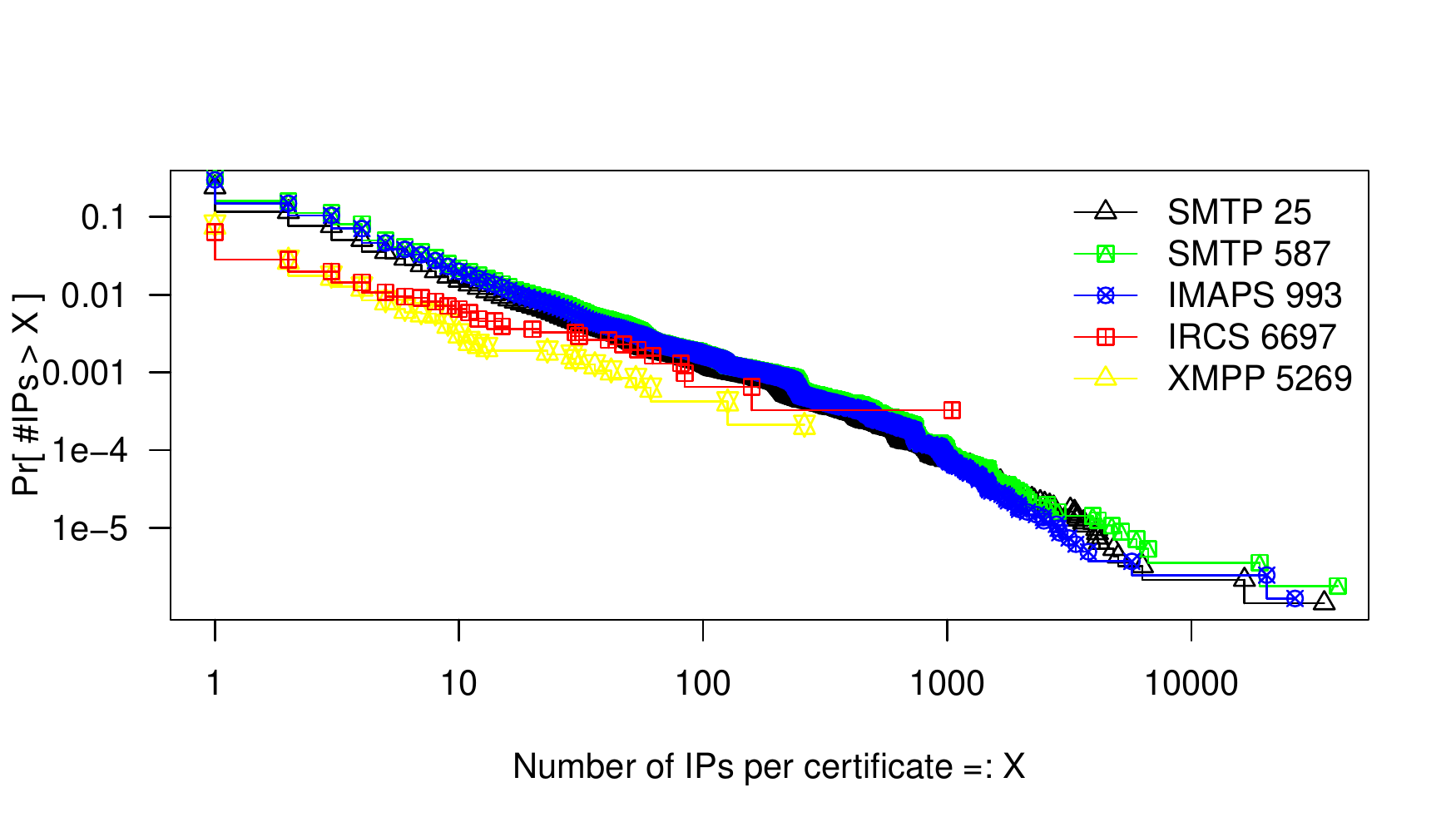}
    \label{fig:cert-occ-selfsigned}
  }

  \caption{Likelihood that a certificate is used on $X$ IPs. \mbox{SMTP 587} is SUBMISSION.}
  \label{fig:cert-occ}
\end{figure}

Holz \etal~\cite{Holz2011} showed that certificates are often reused on
different IP addresses. Although IP addresses do not equal actual hosts, the
frequency at which this phenomenon occurred provided strong indications that
reuse across machines was happening. We investigated this phenomenon here, too.
One potential reason for certificate reuse are Content Distribution Networks
(CNDs). This is a legitimate use case where the ease of key distribution has to
be balanced against a slightly increased attack surface. One would expect a
clear difference in the distributions for valid and invalid certificate chains
in this case as CDNs can be assumed to exercise care in deployment.  Another
possibility are default certificates, potentially from software bundles or
deployed by management tools, which are not changed when the server is further
configured.

\figurename~\ref{fig:cert-occ-all} and~\ref{fig:cert-occ-valid} plot the
likelihood that `a certificate occurs on $X$ IPs' for the entire set of
certificates and only for the set of certificates that have correct certificate
chains, respectively.  While the results for the Web PKI~\cite{Holz2011}
revealed a clear difference between the subset of certificates with valid
chains and the overall set of certificates, this is much less pronounced here.
Furthermore, the likelihood is almost the same across the email protocols.
The only real difference can be seen for XMPP and IRC---however, we need to
stress the smaller number of verifiable certificate chains we have for these
two protocols.

We also investigated the reuse of self-signed certificates. If these are created
purposefully for a single server or service, they should not occur on too many
hosts. Figure \ref{fig:cert-occ-selfsigned} shows, however, that many  appear
on hundreds or thousands of hosts.  Hence, a more likely explanation is that
these are default certificates shipped with software.

\begin{figure}[t!]
  \centering
  \includegraphics[width=0.5\textwidth]{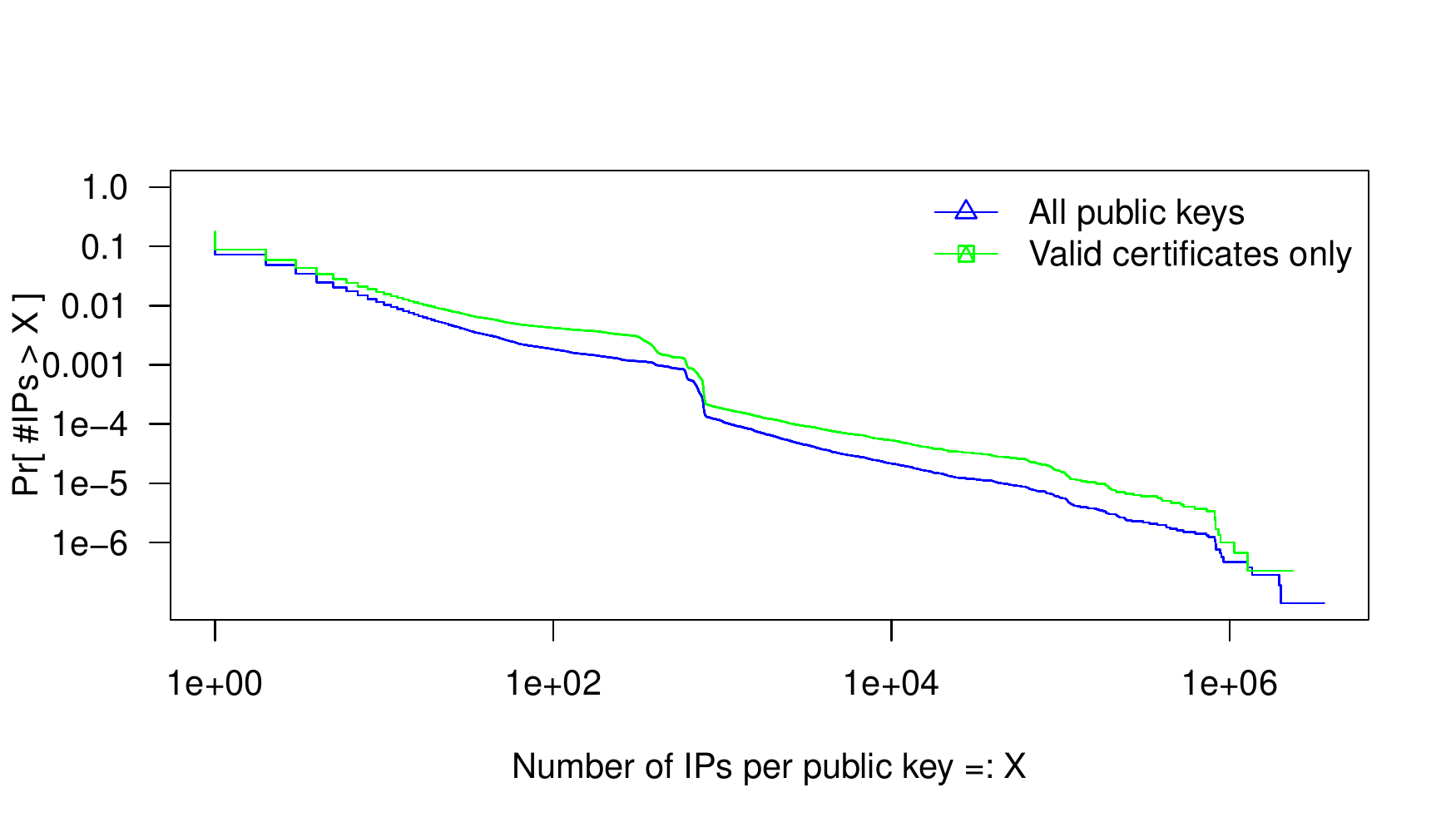}
  \caption{Likelihood that a public key is used on $X$ IPs, across all hosts and certificates.}
  \label{fig:pubkey-occ-all}
\end{figure}

The reuse of certificates is naturally reflected in the number of public keys
that are unique to a host, shown in Figure \ref{fig:pubkey-occ-all}.
Only about 15\% of public keys occur on exactly one host.

\paragraph{Popularity of servers reusing cryptographic material}

We investigated whether passive monitoring would yield similar results for key
reuse. We expected a very different picture as we assume Internet users to
mostly access services of larger providers, which are much more likely to use
correctly deployed certificate chains.

\begin{table}
  \caption{Duplicate certificates by port in passive scans. Entries~marked~with $\dag$ used \starttls.}
  \label{tab:duplicatepassive}
  \begin{center}
    \begin{tabular}{lrrr}
    \toprule
    \multicolumn{1}{c}{\textbf{Protocol}}& \multicolumn{1}{c}{\textbf{Port}}	& \multicolumn{1}{c}{\textbf{Dup. Certs}}	& \multicolumn{1}{c}{\textbf{Valid Dup. Certs}}	\\
    \midrule
    SMTP$^\dag$				& 25					& \num{877}					& \num{656}					\\
    SMTPS				& 465					& \num{36}					& \num{36}					\\
    SUBMISSION$^\dag$			& 587					& \num{46}					& \num{46}					\\
    IMAP$^\dag$				& 143					& \num{29}					& \num{28}					\\
    IMAPS				& 993					& \num{119}					& \num{111}					\\
    POP3$^\dag$				& 110					& \num{12}					& \num{12}					\\
    POP3S				& 995					& \num{43}					& \num{41}					\\
    \midrule
    IRCS				& 6697					& \num{3}					& \num{3}					\\
    XMPP, C2S$^\dag$			& 5222					& \num{35}					& \num{0}					\\
    \bottomrule
    \end{tabular}
  \end{center}
\end{table}

In our passive monitoring run, 1,096 (17\%) of our 6,398 encountered
certificates were seen on more than one IP address.
Table~\ref{tab:duplicatepassive} shows the prevalence of certificate reuses per
port. As the table shows, the majority of certificate reuses happens on port
25.

Furthermore, in our passive scans 78\% of all certificates that we see on at
least 2 hosts are valid, hinting towards the fact that many hosting
providers use this for load balancing. Indeed, examining the certificates that
were seen on the most IP addresses show a SMTP certificate by Proofpoint, Inc.
that was encountered on 263 IPs, followed by Google certificates for
\texttt{imap.gmail.com} (184) and \texttt{mx.google.com} (161).

This shows that, while there is a rampant amount of certificate reuse on the
Internet as a whole, many of these servers seem not to be contacted commonly by
clients, hinting at a considerable server population that might be for private
use or only used by a small user population.

\paragraph{Common names}
\begin{table}[t]
  \caption{Common names in particularly frequently occurring and \textit{invalid} certificates for SMTP, IMAP, XMPP. $\dag$ indicates data obtained during a \starttls negotiation.}
  \label{tab:cert-occ-cn-all}
  \begin{center}
    \begin{tabular}{lr}
      \toprule
      \multicolumn{1}{c}{\textbf{Common name}}		& \multicolumn{1}{c}{\textbf{Occurrences}}	\\
      \midrule
      \multicolumn{2}{c}{\textbf{SMTP$^\dag$}}								\\
      \midrule
      localhost/emailAddress=webaster@localhost		& 35k						\\
      *.bizmw.com					& 34k						\\
      localhost(*)					& 17k						\\
      localhost/emailAddress=webaster@localhost		& 16k						\\
      localhost/emailAddress=webaster@localhost		& 6k						\\
      localhost(*)					& 5k						\\
      plesk/emailAddress=info@plesk.com			& 5k						\\
      localhost/emailAddress=webaster@localhost		& 5k						\\
      localhost(*)					& 4k						\\
      localhost/emailAddress=webaster@localhost		& 4k						\\
      \midrule
      \multicolumn{2}{c}{\textbf{IMAPS}}								\\
      \midrule
      *.securesites.com					& 88k						\\
      *.sslcert35.com					& 31k						\\
      localhost/emailAddress=webaster@localhost		& 27k						\\
      localhost/emailAddress=webaster@localhost		& 21k						\\
      *.he.net						& 19k						\\
      www.update.microsoft.com				& 19k						\\
      *.securesites.net					& 11k						\\
      *.cbeyondhosting2.com				& 11k						\\
      *.hostingterra.com				& 11k						\\
      plesk/emailAddress=info@plesk.com			& 6k						\\
      \midrule
      \multicolumn{2}{c}{\textbf{XMPP, C2S$^\dag$}}							\\
      \midrule
      onex						& 2k						\\
      s2548.pbxtra.fonality.com				& 2k						\\
      k66.ru/emailAddress=postmaster@k66.ru		& 500						\\
      hub.clickmyheart.net				& 400						\\
      John Doe						& 400						\\
      java2go						& 300						\\
      localhost						& 200						\\
      nt-home.ipworldcom				& 200						\\
      mail.visn.net/emailAddress=postmaster@mail.visn.net& 200						\\
      cic-la-plata					& 200						\\
      \bottomrule
    \end{tabular}
  \end{center}
\end{table}

We show the Common Names in some particularly common and invalid certificates
in \tablename~\ref{tab:cert-occ-cn-all}.

Note that we cannot study if the subjects in certificates match the host names
where the certificates are deployed. This would require scans based on a target
list of domain names and comparing the subjects in the received certificates
with the expected domain name. However, we only scanned by IP addresses.
Reverse DNS lookups could theoretically produce domain names to compare
against; however, due to the way that servers are operated today, there is a
risk that the reverse lookup yields hostname aliases different from the actual
domain name by which the server is typically addressed.

Some interesting findings for SMTP on port 25 are as follows.  The certificates
for \texttt{*.bizmw} contain the string `NTT Communications Corporation' in the
`Organisation' part of the subject, a hint in which organisation these invalid
certificates are used. The certificates for `localhost' that are marked with an
asterisk all contain the string `Qmail Toaster Server', thus indicating that
the responsible SMTP server was the popular Qmail by djb.  Presumably, the
operators had never bothered to install proper certificates.  The `webaster'
certificate had already made an appearance in the Web PKI study~\cite{Holz2011}
and is most likely due to a certificate creation software with a spelling
weakness. Plesk is the company behind the Parallels visualisation product.

For IMAPS, we also find the popular `webaster' certificate. Furthermore we find
certificates of several hosting companies and also of Hurricane Electric. The
certificates for Microsoft were a surprise as they seemed to contain a Web
address for the Windows Update service. There were \num{18193} occurrences of
this single end-host certificate. No intermediate or root certificates were
sent. The respective hosts were distributed across 15 Autonomous Systems, which
we looked up using the Team Cymru ASN
Database\footnote{\texttt{https://asn.cymru.com/cgi-bin/whois.cgi}}.

\begin{table}[t]
  \caption{Invalid Microsoft certificates: ASes and CIRCL ranking for botnet and malicious activity.}
  \label{tab:mscerts}
  \begin{center}
    \begin{tabular}{lp{4.5cm}r}
        \toprule
	\multicolumn{1}{c}{\textbf{AS number}}& \multicolumn{1}{c}{\textbf{Registration information}}& \multicolumn{1}{c}{\textbf{CIRCL rank}} \\
	\midrule
	3257			& TINET-BACKBONE Tinet SpA, DE			& 9532		\\
	3731			& AFNCA-ASN - AFNCA Inc., US			& 4804		\\
	4250			& ALENT-ASN-1 - Alentus Corporation, US		& 9180		\\
	4436			& AS-GTT-4436 - nLayer Communications, Inc., US & \num{10730}	\\
	6762			& SEABONE-NET TELECOM ITALIA SPARKLE S.p.A., IT & \num{11887}	\\
	11346			& CIAS - Critical Issue Inc., US		& 557		\\
	13030			& INIT7 Init7 (Switzerland) Ltd., CH		& 6255		\\
	14618			& Amazon.com Inc., US				& 4139		\\
	16509			& Amazon.com Inc., US				& 3143		\\
	18779			& EGIHOSTING - EGIHosting, US			& 4712		\\
	21321			& ARETI-AS Areti Internet Ltd.,GB		& 2828		\\
	23352			& SERVERCENTRAL - Server Central Network, US	& \num{11135}	\\
	26642			& AFAS - AnchorFree Inc., US			& --		\\
	41095			& IPTP IPTP LTD, NL				& 6330		\\
	54500			& 18779 - EGIHosting, US			& --		\\
        \bottomrule
        \end{tabular}
  \end{center}
\end{table}

\tablename~\ref{tab:mscerts} shows the results of the lookups. None of the
ASes were registered to Microsoft; they were predominantly assigned to
hosters. We checked the BGP ranking of these ASes with CIRCL's web site, which
as of 12 August 2015 contained \num{12339} ASes ranked for known botnet and
malicious activity. Only two of the ASes were not on that list. Manual
inspection of the certificate did not yield anything out of the ordinary,
however. We contacted Microsoft repeatedly (directly and via CIRCL), but never
received a response why such a certificate should occur on an email port.

Analyzing XMPP certificates also yielded some interesting results. OneX is an
XMPP server by Avaya, a communications company---this seems to be a default
certificate.  Fonality is a provider of unified messaging. The certificates for
\texttt{k66.ru} contained a string referring to a product called `CommuniGate'. It 
also appeared in the certificates for \texttt{visn.net}. We were unable to
determine the exact nature of \texttt{clickmyheart}, but the Web site shows a
login portal, so we presume some forum. The certificates also contained the
name Zimbra Collaboration server. `John Doe' is used in the certificates by
Jive Software.  Java2go seems to be an SMS product.

Beyond the strange Microsoft Update certificate, our results suggest that a
number of default keys and certificates are used in production.  This is a
negative finding as it means that other parties may have access to the private
cryptographic material.  Some vendors, meanwhile, seem to choose their own, private CA
instead of working with a commercial one.

\subsection{Poor CA practice}
\label{sub:poorcapractice}

In our data set, we were still able to find certificates that were issued
directly from a root CA without any intermediate certificate. The industry has
moved away from this practice and discourages it today \cite{CABForumBR2011}.
To issue such certificates, the CA's root certificate needs to be kept online,
a serious attack vector. A root certificate
compromise would necessitate an update of all clients that include it. We
expect the number of certificates that are directly issued from a root
certificate to shrink further.

Indeed, there were very few cases already in our scan. For SMTP, for instance,
we found only 794 cases, or 0.07\% of verifiable chains. The percentages for
SUBMISSION was 0.08\%. Interestingly, it was 0.5\% for SMTPS.  SMTPS is
deprecated---it is not implausible that operators who still enable SMTPS have
simply never upgraded to new, intermediate-issued certificates.  For good
measure, we also tested this property for the two IRC protocols (only one case
for IRCS), the two XMPP client-to-server variants (two for STARTTLS, 14 for
XMPPS), and the two XMPP server-to-server variants (one case each).

\subsection{Authentication methods}

\begin{table}[t]
  \caption{Authentication mechanisms offered by servers on SUBMISSION port.}
  \label{tab:submission_auth_unique}
  \begin{center}
    \begin{tabular}{lrrr}
    \toprule
    \multicolumn{1}{c}{\textbf{Mechanism}}	& \multicolumn{1}{c}{\textbf{Advertised}}	& \multicolumn{1}{c}{\textbf{Servers}} \\
    \midrule
    PLAIN					& \num{2764157}					& 99.27\%					\\
    LOGIN					& \num{2760100}					& 99.15\%					\\
    CRAM-MD5					& \num{431634}					& 15.50\%					\\
    DIGEST-MD5					& \num{230152}					& 8.26\%					\\
    OTP						& \num{19850}					& 0.71\%					\\
    GSSAPI					& \num{16555}					& 0.59\%					\\
    NTLM					& \num{13663}					& 0.49\%					\\
    XOAUTH2					& \num{3118}					& 0.11\%					\\
    PLAIN-CLIENTTOKEN				& \num{1642}					& 0.05\%					\\
    XOAUTH					& \num{1641}					& 0.05\%					\\
    \midrule
    Other \num{591} mechanisms found		& \num{5329}					& 0.19\%					\\
    \bottomrule
    \end{tabular}
  \end{center}
\end{table}

\begin{table}[t]
  \caption{Authentication mechanisms offered by servers on IMAPS port.}
  \label{tab:submission_auth_unique_imaps}
  \begin{center}
    \begin{tabular}{lrrr}
    \toprule
    \multicolumn{1}{c}{\textbf{Mechanism}}	& \multicolumn{1}{c}{\textbf{Advertised}}	& \multicolumn{1}{c}{\textbf{Servers}} \\
    \midrule
    PLAIN					& \num{3753658}					& 96.66\%					\\
    LOGIN					& \num{2430559}					& 62.59\%					\\
    CRAM-MD5					& \num{467460}					& 12.04\%					\\
    CRAM-SHA1					& \num{186355}					& 4.80\%					\\
    CRAM-SHA256							& \num{185427}					& 4.77\%					\\
    DIGEST-MD5					& \num{160893}					& 4.14\%					\\
    GSSAPI					& \num{18851}					& 0.49\%					\\
    NTLM					& \num{17106}					& 0.44\%					\\
    X-ZIMBRA				& \num{7582}					& 0.20\%					\\
    MSN					& \num{4181}					& 0.11\%					\\
    \midrule
    Other \num{61} mechanisms found		& \num{6773}					& 0.17\%					\\
    \bottomrule
    \end{tabular}
  \end{center}
\end{table}

We analysed the authentication mechanisms supported and advertised by servers
to clients when sending mails using SUBMISSION or retrieving mails with IMAPS.
In our active measurements, we queried the servers for authentication
capabilities using the EHLO command for SUBMISSION and the CAPABILITIES command
for IMAPS.  Capabilities were always queried after TLS session establishment.
We show the most common authentication mechanisms in
Tables~\ref{tab:submission_auth_unique}
and~\ref{tab:submission_auth_unique_imaps}.

The results obtained for both SUBMISSION and IMAPS show poor support for strong
authentication mechanisms. Mechanisms transmitting credentials in plaintext
(PLAIN and LOGIN) are supported by more than 99\% of the SUBMISSION and 90\% of
the IMAPS servers. On the other hand, less than 16\% of the SUBMISSION and
12.04\% of the IMAPS servers support much stronger mechanisms such as CRAM.

\begin{table}[t]
  \caption{Combinations of authentication mechanisms offered by servers on SUBMISSION port\label{tab:submission_auth_combinations}.}
  \begin{center}
    \begin{tabular}{p{5.3cm}rrr}
    \toprule
    \multicolumn{1}{c}{\textbf{Mechanism}}	& \multicolumn{1}{c}{\textbf{Advertised}}	& \multicolumn{1}{c}{\textbf{Servers}} \\
    \midrule
    PLAIN, LOGIN				& \num{2092594}					& 75.15\%					\\
    LOGIN, PLAIN				& \num{224197}					& 8.51\%					\\
    LOGIN, CRAM-MD5, PLAIN			& \num{96322}					& 3.45\%					\\
    LOGIN, PLAIN, CRAM-MD5			& \num{45477}					& 1.63\%					\\
    DIGEST-MD5, CRAM-MD5, PLAIN, LOGIN		& \num{36416}					& 1.30\%					\\
    CRAM-MD5, PLAIN, LOGIN			& \num{29046}					& 1.04\%					\\
    PLAIN, LOGIN, CRAM-MD5			& \num{24914}					& 0.89\%					\\
    CRAM-MD5, DIGEST-MD5, LOGIN, PLAIN		& \num{19877}					& 0.71\%					\\
    PLAIN					& \num{17079}					& 0.61\%					\\
    \midrule
    Other \num{1234} combinations found		& \num{326392}					& 7.11\%					\\
    \bottomrule
    \end{tabular}
  \end{center}
\end{table}

\begin{table}[t]
  \caption{Combinations of authentication mechanisms offered by servers on IMAPS port\label{tab:submission_auth_combinations_imaps}.}
  \begin{center}
    \begin{tabular}{p{5.3cm}rrr}
    \toprule
    \multicolumn{1}{c}{\textbf{Mechanism}}	& \multicolumn{1}{c}{\textbf{Advertised}}	& \multicolumn{1}{c}{\textbf{Servers}} \\
    \midrule
    PLAIN, LOGIN	                              & \num{2222721}			& 60.16\%	\\
    PLAIN				                                & \num{982386}			& 26.59\% \\
    CRAM-MD5, CRAM-SHA1, CRAM-SHA256, PLAIN			& \num{183813}			& 4.97\%  \\
    CRAM-MD5, PLAIN                             & \num{90341}			  & 2.45\%  \\
    PLAIN, LOGIN, DIGEST-MD5, CRAM-MD5          & \num{78061}				& 2.11\%	\\
    LOGIN                                       & \num{21842}					& 0.59\%\\
    CRAM-MD5, PLAIN, LOGIN, DIGEST-MD5          & \num{16660}					& 0.45\%\\
    PLAIN, LOGIN, CRAM-MD5                      & \num{10731}				& 0.29\%	\\
    CRAM-MD5, PLAIN, LOGIN, DIGEST-MD5, NTLM		& \num{9105}					& 0.25\%\\
    PLAIN, X-ZIMBRA		                          & \num{7569}					& 0.20\%\\
    \midrule
    Other \num{1039} combinations found		& \num{71685}					& 1.94\%		\\
    \bottomrule
    \end{tabular}
  \end{center}
\end{table}

This is made worse by the fact that the vast majority of SUBMISSION (84.86\%)
and IMAPS (87.43\%) servers support \emph{only} PLAIN and LOGIN.  The ordering
of authentication mechanisms is not particularly encouraging, either: clients
obeying the ordering suggested by many servers will use a plaintext mechanism
for SUBMISSION (resp. IMAPS) in at least 96.19\% (resp. 89.35\%) of the cases
(Tables~\ref{tab:submission_auth_combinations}
and~\ref{tab:submission_auth_combinations_imaps})

\begin{table}[t]
  \caption{Authentication mechanisms observed in connections on SUBMISSION port.}
  \label{tab:submission_auth_passive}
  \begin{center}
    \begin{tabular}{lrrr}
    \toprule
    \multicolumn{1}{c}{\textbf{Mechanism}}	&
		\multicolumn{1}{c}{\textbf{Connections}}	& \multicolumn{1}{c}{\textbf{Servers}} \\
    \midrule
    PLAIN					& 4.4\% & 39\%	\\
    LOGIN				  & 4.3\% & 37\%	\\
    CRAM-MD5      & 0.7\% & 10\%	\\
    DIGEST-MD5		&	0.5\% & 3.7\%	\\
		XAOL-UAS-MB   & 0.4\% & 1.8\%\\
		GSSAPI        & 0.3\% & 4.3\% \\
    NTLM					&	0.3\% & 3.7\% \\
    XOAUTH2				&	0.03\% & 1.2\%\\
    XYMCOOKIE			&	0.01\% & 0.6\%\\
    \bottomrule
    \end{tabular}
  \end{center}
\end{table}

\begin{table}[t]
  \caption{Combinations of authentication mechanisms observed in connections on SUBMISSION port.\label{tab:submission_auth_combinations_passive}.}
  \begin{center}
    \begin{tabular}{p{5.2cm}rrr}
    \toprule
    \multicolumn{1}{c}{\textbf{Mechanisms}}	&
		\multicolumn{1}{c}{\textbf{Connections}}	& \multicolumn{1}{c}{\textbf{Servers}} \\
    \midrule
    PLAIN, LOGIN				& 1.82\%				& 20.86\%					\\
    LOGIN, PLAIN				& 1.68\%				& 18.40\%					\\
    LOGIN,PLAIN,XAOL-UAS-MB			& 0.19\% & 1.84\%	\\
    PLAIN,LOGIN,XAOL-UAS-MB			& 0.16\% & 1.84\%	\\
    GSSAPI		& 0.11\% & 1.23\%	\\
    GSSAPI,NTLM		& 0.10\% & 1.84\%	\\
    LOGIN,PLAIN,CRAM-MD5			& 0.09\% & 3.68\%	\\
    DIGEST-MD5,CRAM-MD5		& 0.09\%  & 0.61\%	\\
    CRAM-MD5,DIGEST-MD5	& 0.09\% & 0.61\%	\\
    PLAIN,LOGIN,CRAM-MD5 & 0.08\% & 0.61\%	\\
    \midrule
    Other \num{17} combinations observed	& 1.09\%	& 14.68\% \\
    \bottomrule
    \end{tabular}
  \end{center}
\end{table}

In addition, as part of our passive data collection, we also measured which
authentication methods servers offered. In contrast to our active measurement,
we can only record the authentication methods offered \emph{before} encryption
starts, not those after encryption has started.
Table~\ref{tab:submission_auth_passive} shows the percentage of servers that
offer a certain authentication mechanism as well as the percentage of
connections in which a certain authentication mechanism was offered.
Table~\ref{tab:submission_auth_combinations_passive} shows the combination of
authentication mechanisms offered that we observed.

The results here are not encouraging---while, according to
Table~\ref{tab:submission_auth_combinations_passive}, only 68.88\% of servers
offer authentication before STARTTLS, 39.87\% of all servers offer only
 authentication based on PLAIN and LOGIN. When looking at the
number of connections, this picture is even more pronounced with only 4.94\% of
connections containing information about authentication mechanisms before
STARTTLS, but 3.51\% of all observed connections containing only plaintext
authentication mechanisms. This is consistent with our findings in
Section~\ref{sec:starttls} that showed that 97\% of SUBMISSION connections
upgraded their connection using STARTTLS.  Nonetheless, 71\% of clients who did not
upgrade their connections also used plaintext mechanisms to
authenticate.

Moreover, 31\% of observed IMAP servers (serving 16\% of passively observed
connections) refused plaintext logins before encryption (with the
LOGINDISABLED capability).

\section{Risks, threats, and mitigation}
\label{sec:risksthreats}

In this section, we first summarise the current risks and threats to Internet
communication protocols, based on our analyses.  We also present
recommendations on how to improve the situation in the future.  For our
discussion, it is important to consider which attacker one wishes to defend
against: some security configurations are strong enough against passive,
eavesdropping attackers.  They are thus secure against global pervasive
monitoring of traffic: active attacks require much higher effort and can, in
all likelihood, not be carried out on global scale yet.

\subsection{\starttls semantics}

Many of the discussed communication protocols, and especially SMTP, rely on
\starttls to initiate encrypted connections. The problem is that, as shown in
Section~\ref{sec:starttls}, less than 51\% of servers support upgrading
connections to \tls.  Fortunately, some providers pushed strongly for better
adoption of \tls in the last years, increasing the share of connections that
use \tls by a significant amount.  Nonetheless, in the interest of reliability,
many clients and servers will fall back to non-encrypted connections should
\starttls not be offered.

\subsection{Cipher use}

A common problem is the continued choice of weak ciphers in communication
protocols. Depending on the protocol, we still see up to 17\% of connections
choosing RC4, which has been deprecated in~\cite{rfc7465}. This is in contrast
to the Web at large, where currently about 10\% of connections use RC4 ciphers,
according to the ICSI Certificate Notary.
This difference is likely caused by the push to increment Web security in the
last few years, where academia, industry, and the open source community have
driven adoption of more modern ciphers. This movement does not seem to have
reached other, non-web protocols. Another fact that supports this hypothesis is
the high number of connections that do not use forward-secure ciphers (more
than 30\% for some of the protocols we observed).

Another issue is the use of Diffie-Hellman parameter sizes of 1024 bits or less
in more than $\frac{2}{3}$ of all observed connections.  While there are some
limited cases in which this might be necessary for legacy
compatibility\footnote{Java$\le$7 cannot use DH parameters larger than 1024
bits; see
\texttt{http://blog.ivanristic.com/2014/03/ssl-tls-improvements\\-in-java-8.html}.},
it seems unlikely that this is a conscious choice by server operators.

\subsection{Certificate chains and their validity}

We showed that a high number of servers, especially as compared to earlier
scans of the HTTPS protocol, serve chains using broken or self-signed
certificates.  This is of particular importance for SMTP with STARTTLS.  If
servers with unverifiable certificates are actually able to receive email from
\ssltls-capable servers, this suggests that a significant number of servers are 
not verifying certificates in outgoing connections or ignore certificate
errors.  This means that there are likely many cases where \ssltls does not
provide \pitm protection in \stos connections, a finding that is also supported
by \cite{Foster2015}. As noted before, SMTP tends to
prioritise delivery before security (for good reason), and operators will
likely favour such an approach to ensure messages reach their destination.
On a more positive note, our passive data did reveal that, unlike the servers at
large, a much larger ratio of the connections we observed used chains that we
could verify. We assume this to be due to the fact that large providers are more
capable and willing to invest the time in securing their \ssltls setups.

\subsection{Authentication methods}

The more secure challenge-response authentication mechanisms for SMTP and IMAP do
not seem to enjoy much popularity: the methods PLAIN and LOGIN are preferred.
We speculate this is because operators tend to assume their connections are
already secure because of the use of \ssltls---however, this is only true if
certificates are actually correct and strong ciphers are used.  We also found
that client authentication methods are sometimes offered before the
negotiation of a \ssltls session---the vast majority being PLAIN and LOGIN again.
This means that an eavesdropper or active \pitm can collect the plaintext of
specific messages and is also able to acquire login credentials.

\subsection{Suggestions for improvement}
\setcounter{paragraph}{0}

We offer several actionable recommendations based on our analyses.

\paragraph{Observable infrastructure}

One key element to improving messaging security is to create awareness. We
believe better observability can be provided in a two-pronged approach: through
more regular active scans, but also by facilitating observation of the security
of one's own communication.  Much insight can be gained by security-conscious
users by information available on the dashboards mentioned in
Section~\ref{sec:relatedwork}.  We believe that current efforts like SSL Pulse,
which aims to enhance the use and security of \tls on the Web, should be
extended to other communication protocols. The coverage should also be extended
to the entire Internet. The data provided by \texttt{censys.io}
\cite{Durumeric2015a} is a good source. Auditing mechanisms that so far are
meant primarily for use on the Web, such as Certificate Transparency by Google,
could also be extended to include email or chat servers.

Another way to improve the situation would be for client software to have
clearer user warnings when connecting to servers in insecure ways.  Unlike the
Web, the messaging infrastructure uses intermediate relays. It is, however,
possible to derive (partial) information about the path an email has taken or
is going to take \cite{Foster2015}. User-agents could be extended to report
information about this and signal if the used cryptography is deemed secure.

\paragraph{Deployability and configuration}

The high number of invalid or otherwise unverifiable certificate chains is a serious obstacle
to the ubiquitous adoption of STARTTLS for email forwarding. Two of the reasons
for this poor deployment are the costs of certification by CAs and the
high difficulty of proper server configuration. The Let's
Encrypt\footnote{\texttt{https://letsencrypt.org/}} initiative, which currently
focuses on Web certification, addresses both issues and should be extended to
include email and chat. However, configuration complexity is not straight-forward to
address. BetterCrypto.org\footnote{\texttt{https://bettercrypto.org}}, for example,
provides a guide to correct \tls configuration. This approach scales poorly
as almost every software suite uses its own configuration format (or so it
often seems). We thus call for a standard for \emph{unified \ssltls
configuration}: there is no reason why certificates and cryptography cannot be
configured using a unified syntax. Configuration files for \ssltls could simply
be included (or parsed) in the general configuration of a software.

A recent RFC \cite{rfc7672} describes how to increase the security of SMTP
opportunistically by using the TLSA record of DNS~\cite{rfc6698} to signal \tls
support. This removes CAs as a single point of failure.
Due to its opportunistic nature, the approach can be deployed
incrementally. It defeats attackers carrying out global pervasive monitoring.
However, DNS/DNSSEC is once again a complex system---thus, this approach also
calls for better configuration support and a unified configuration syntax.

\paragraph{Flag-days for mandatory encryption}

Large providers should research the impact of refusing insecure
connections---this may cause smaller providers to use verifiable chains.  To
preserve reliability of delivery, providers could deploy a form of grey-listing for
senders using insecure connections and only offer fast message delivery to mail
servers capable of using \ssltls in a secure fashion.

A broad call for mandatory encryption and enforcing \mbox{\starttls} before
using a connection could further push the adoption of better security
practices. An example showing that community-based actions have a chance of
succeeding is
XMPP\footnote{\texttt{http://blog.prosody.im/mandatory-encryption-on-xmpp\\-starts-today/}}.
This can be combined with a pinning-like approach where clients
refuse plaintext connections, especially to popular servers, if previous
connections had used STARTTLS. This is in line with the findings
of~\cite{2015durumeric_empirical_analysis_mail_security}, where the authors
show that certain ISPs may try to downgrade connections by stripping STARTTLS
commands.

A further step to foster the proper use of \ssltls could be taken by package
and operating system maintainers. If packages shipped with a safe, modern
configuration for the ciphers to use, many of the problems we highlighted could
be remedied. A summary of the best current practices for this purpose can be
found in~\cite{rfc7525}.  Conversely, application packages should never
be shipped with default cryptographic material, not even as examples.  A saner
approach would be to provide scripts to generate the keys and certificates.

\paragraph{Application-layer authentication}

Challenge-response authentication mechanisms like CRAM and SCRAM avoid sending
credentials over the wire. SCRAM even allows servers to store the password in
non-plaintext form, thus combining improved credential storage with safer
authentication. Support for challenge-response forms of authentication seems to
be lacking, possibly because developers see no need for it and operators fear the
complexity. Standard packages and unified configuration can help here.

\section{Conclusion}
\label{sec:conclusion}

This paper presented the largest study of the security of the standard Internet
messaging infrastructure to date. Based on active scans of servers and passive
monitoring of client connections, we collected the parameters used to establish
\ssltls sessions, the details of X.509 certificates offered by servers, and the
application-layer authentication methods offered to, and used by, clients.
Across the whole Internet, we found a worryingly high number of poorly secured
servers. This was either due to cryptographic parameter and cipher choices or
due to invalid or duplicated cryptographic material. Too many servers also
offer weak application-layer authentication methods. A silver lining  is that
there are significantly better deployments in the most popular services, and a
majority of observed clients connected using reasonably secure parameters when
they did request encryption.  Nonetheless, too many of the connections we
observed were still performed in the clear. Moreover, we found that many \ctos
setups, especially for SMTP, did not use valid credentials. This means that
email in transit may often be delivered over unencrypted and unauthenticated
hops. We gave a list of recommendations that are actionable and can help
to significantly improve the situation.

\section*{Acknowledgments}

This work was supported by the National Science Foundation under grant numbers
CNS-1528156 and ACI-1348077. Any opinions, findings, and conclusions or
recommendations expressed in this material are those of the author(s) and do
not necessarily reflect the views of the NSF.

\renewcommand{\bibfont}{\small}
\printbibliography

\end{document}